\documentclass[nofootinbib]{revtex4}
\usepackage{graphicx}
\usepackage{amssymb}
\usepackage{amsmath}
\usepackage{bm}
\usepackage{hyperref}%for .pdf
\begin{document}

\title{Growth of structure in interacting vacuum cosmologies}
\author{Humberto A.~Borges$^{1,}$$^{2}$, David Wands$^{1}$}

\affiliation{$^1$Institute of Cosmology and Gravitation, University of
Portsmouth, Dennis Sciama Building, Portsmouth, PO1 3FX, United Kingdom\\$^2$Instituto de F\'isica, Universidade Federal da Bahia, Salvador, BA, 40210-340, Brasil}

 \date{7th April 2020}

\begin{abstract}
We examine the growth of structure in three different cosmological models with interacting dark matter and vacuum energy. We consider the case of geodesic dark matter with zero sound speed, where the relativistic growing mode in comoving-synchronous gauge coincides with the Newtonian growing mode at first order in $\Lambda$CDM. We study corrections to the linearly growing mode in the presence of interactions and the linear matter growth rate, $f_1$, contrasting this with the velocity divergence, $f_{\rm rsd}\sigma_8$, observed through redshift-space distortions. We then derive second-order density perturbations in these interacting models. We identify the reduced bispectrum that corresponds to the non-linear growth of structure and show how the shape of the bispectrum is altered by energy transfer to or from the vacuum. Thus the bispectrum, or higher-order correlators, might in future be used to identify dark matter interactions.
\end{abstract} 

\maketitle

\section{Introduction}\label{intro}

The current accelerated expansion of the universe, inferred from observations of type Ia supernovae (SNe Ia) \cite{Astier,riess,perl}, anisotropies in the cosmic microwave background (CMB) and observations of large-scale structures (LSS), among others, is one of the most fascinating topics in modern cosmology, attracting the attention of researchers in both the theoretical and experimental area. The most common explanation is the existence of an energy component that has negative pressure known as ``dark energy" \cite{tegmark}, which in its simplest form corresponds to a cosmological constant in the Einstein equations of general relativity \cite{pebles,Padmanaban,weinberg02}. Observations show that around $95\%$ of the energy in the Universe today is in the form of dark energy and dark matter, which plays a crucial role in the formation of galaxies and clusters of galaxies.

Cosmology with a cosmological constant and cold dark matter has become the standard model of the universe, known as $\Lambda$CDM. This model has proved to be successful when tested against a range of precise observations \cite{planck}. However, despite these successes, the problem remains that the vacuum energy density observed today is much lower than the theoretical value predicted by quantum field theories \cite{weinberg02}. Thus there is a need to find a mechanism to understand the small value of the dark energy density required by observations. If the origin of dark energy is not a cosmological constant, then alternative models \cite{Ozer} should be considered to explain the current accelerated expansion of the universe. Often this is done by introducing additional fields whose dynamics modify the dark energy equation of state and determine the present density \cite{Copeland:2006wr}.

An alternative approach is to instead consider an interacting vacuum energy whose present value is dependent on energy-momentum transfer with existing matter 
%v2
fields\footnote{This differs from interacting dark energy models which introduce additional dark energy fields interacting with dark matter~\cite{Wetterich:1994bg,Amendola:1999qq,Holden:1999hm,He:2008tn,Valiviita:2008iv,Koyama:2009gd,Tsujikawa:2012hv,Marcondes:2016reb}.
}. Since the physics underlying the dark sector is still unknown, it could be that vacuum energy and dark matter interact directly and exchange energy. Unified dark matter models, such as the generalised Chaplygin gas (gCg) \cite{kam, Fabris, Bento,Sandvik:2002jz}, can easily be decomposed into two interacting components \cite{bento,degeneracy}, one representing dark matter density, $\rho_{\rm dm}$, and the other the vacuum energy, $\rho_V$. The energy exchange implied by this decomposition can be written for the gCg model as $Q=3\alpha H\rho_{\rm dm}\rho_{V}/\rho$ \cite{Wands}, where $\alpha$ is a dimensionless parameter constant. For $\alpha<0$ there is more matter today compared with $\Lambda$CDM if we start with the same amount of primordial matter at high redshift. One particular case is given by $\alpha=-0.5$, which corresponds to a dark matter created at a constant rate due to a decaying vacuum energy \cite{Borges}. This particular model has been shown competitive with the $\Lambda$CDM model when tested against observational data including LSS, SNe Ia and integrated Sachs-Wolfe (ISW) constraints \cite{saulo, hermano}. 
On the other hand a full analysis of CMB+ISW constraints on the decomposed gCg model gives the bounds $-0.15<\alpha<0.26$ \cite{Wang:2013qy}, while a joint analysis of LSS, SNe Ia and the position of the first peak of CMB has lead to $-0.39<\alpha<-0.04$ ($2\sigma$) \cite{note}. The results of analysis using Planck data for the CMB anisotropy spectrum is consistent with $|\alpha| \leq 0.05$ \cite{vom}.

An interaction of the form $Q=-q_VH\rho_V$ \cite{Salvatelli,Martinelli:2019dau} has also been studied in light of observations, with $q$ taking different values in distinct redshift bins. The analyses suggested that a non-zero interaction may be favoured by cosmological data, including redshift-space distortions, when compared with $\Lambda$CDM model. 
Another interaction, proposed in 
% v2
\cite{Shapiro:2003ui,EspanaBonet:2003vk,Wang1},
%in context of decaying vacuum scenario in which dark matter dilute more slowly than $\bar{\rho}_{\rm dm}\propto a^{-3}$. The correspondent interaction between vacuum and matter 
is $Q=\epsilon H\rho_{\rm dm}$ 
%whose deviation from the standard evolution is characterized by 
with a small constant $\epsilon$. Such a scenario is obtained in Ref. \cite{jailson} from thermodynamics arguments. The best fit found is $\epsilon=-0.11$ through a joint analysis involving measurements of type Ia supernovae, gas mass fraction and CMB. Ref.~\cite{valent} found $\epsilon\sim-10^{-2}$,
% v2
and some authors have argued \cite{Sola:2016jky,Sola:2016ecz,Sola:2017jbl} that there is evidence for $\epsilon<0$ at more than 4$\sigma$ including LSS data. 
An approach to construct model-independent constraints on the dark matter-vacuum interaction is presented in \cite{Wang:2015wga,Hogg:2020rdp}.

At the same time, it is widely believed that another period of accelerated expansion called inflation occurred at very high energies in the very early universe and primordial perturbations were created from quantum fluctuations; this creates the seed for large-scale structures that grow by gravitational instability to result in the present distribution of matter on cosmological scales. 
%In the standard inflationary model the seeds for LSS are generated on scales outside the horizon from the flutuations of a scalar field called inflaton.
A non-Gaussian distribution of primordial perturbations, that appears due to nonlinear evolution in second-order perturbation theory, has been proposed as a means to discriminate among different inflationary scenarios. Gravitational instability is a non-linear process which itself leads to non-Gaussianity in the matter distribution at late times, even if we start with a completely Gaussian perturbation. Thus it is important to understand the effects of nonlinear evolution, including possible interactions between vacuum energy and dark matter, in order to be able to distinguish possible non-linear effects of vacuum interactions from those of primordial non-Gaussianity.

In this work we study both linear and non-linear evolution of matter perturbations \cite{matarrese, Noh:2005hc, bartolo, bartolo1, Bruni, Bruni:2014xma, Uggla, Chani} in the presence of an interacting vacuum energy. We employ the fluid-flow approach adopted in \cite{Bruni}, including for the first time the effects of energy transfer in gravitational clustering at second order, as well as making a careful study of peculiar velocities and hence redshift-space distortions in the presence of interactions. At second order we identify the effects of primordial non-Gaussianity and non-linear growth of structure, leading to distinct shapes for the reduced bispectrum at second-order.

\section{Fluid-flow equations}

\label{II}

The Einstein field equations are given by
\begin{equation}\label{um}
R_{\mu\nu}-\frac{1}{2}g_{\mu\nu}R=T_{\mu\nu},
\end{equation}
where $R_{\mu\nu}$ represents the Ricci tensor, R the Ricci scalar, and $g_{\mu\nu}$ represents the space-time metric.
% $T_{\mu\nu}$ is the total energy-momentum tensor of matter and vacuum. 
We will consider pressureless dark matter, $p_{\rm dm}=0$, with energy density $\rho_{\rm dm}$ and vacuum energy, $\rho_V$, with equation of state $p_V=-\rho_V$, such that the energy-momentum tensor of matter plus vacuum is
\begin{equation}
T_{\mu\nu} =  {T_{({\rm dm})\mu\nu}} + {T_{(V)\mu\nu}} = \rho_{\rm dm}u_\mu u_\nu - \rho_V g_{\mu\nu} \,.
\end{equation}
where $u^\mu$ is the matter four-velocity.
The energy-momentum conservation equations for each component are given by
\begin{equation}\label{asdw}
\nabla^\mu {T_{(V)\mu\nu}} = Q_{\nu}\,,
\end{equation}
\begin{equation}\label{asd}
\nabla^\mu {T_{({\rm dm})\mu\nu}} = -Q_{\nu}\,,
\end{equation}
where the energy-momentum transfer from the dark matter to the vacuum is $Q_{\mu}=-\nabla_\mu\rho_V=\nabla_\mu p_V$.

We will assume\footnote{Another possibility, for example, would be that the energy flow follows the gradient of matter density, which implies that the local vacuum energy is a function of the local matter density. In that case the sound speed corresponds to the adiabatic sound speed, as in unified dark matter models with barotropic equation of state, and the energy transfer is already strongly constrained by CMB observations \cite{Sandvik:2002jz}.} that the energy transfer follows the 4-velocity of the dark matter, $Q^\mu=Qu^\mu$ \cite{Wands}. This has two important consequences. Firstly, the vacuum is homogeneous on hypersurfaces orthogonal to the matter 4-velocity. This means that there are no pressure gradients in a frame comoving with matter. Thus matter follows geodesics and the matter sound speed is zero. Secondly, the matter 4-velocity is a potential flow and thus irrotational. We expect this to be a good description of matter at early times and on large scales where the initial density field is set by primordial scalar perturbations. This is sufficient for our perturbative treatment of the initial growth of structure, but at late times we would expect the nonlinear growth of structures to develop vorticity and indeed to develop rotationally supported dark matter halos. Thus we expect the geodesic approximation to break down below some length scale. Otherwise truly irrotational dark matter would have distinctive observational consequences \cite{Sawicki:2013wja}.

Since there are no pressure gradients orthogonal to the matter 4-velocity, we can write the equations of motion in a comoving-synchronous gauge,
just as in $\Lambda$CDM,
%, where each observer has 4-velocity $u_{\mu}=(-a,0,0,0)$ and a universal time. 
%In this gauge the vacuum not clustered. Therefore, the metric spatially flat for the comoving synchronous gauge is
where we write the line element as
\begin{equation}
ds^2=a^2(\eta)[-d\eta^2+\gamma_{ij}dx^idx^j] \,.
\end{equation}
%and $\gamma_{ij}$ is the comoving spatial metric. 

We will consider inhomogeneous perturbations about a spatially flat Friedmann-Robertson-Walker background for which $\bar\gamma_{ij}=\delta_{ij}$ and we use an overbar to denote the spatially homogeneous background solution. The background expansion is given by the Friedmann constraint equation
\begin{equation}\label{aore}
3\mathcal{H}^2=a^2(\bar\rho_{\rm dm}+\bar\rho_V)
%=a^2\bar\rho
\,,
\end{equation}
where the conformal Hubble rate is $\mathcal{H}\equiv a'/a$ and a prime denotes a derivative with respect to conformal time.

Following \cite{Bruni,bartolo}, we define the deformation tensor by the conformal time derivative of the spatial metric
\begin{equation}\label{ab}
\vartheta^i_j=\frac{1}{2}\gamma^{ik}\gamma'_{jk} \,,
\end{equation}
and the perturbed scalar expansion by
\begin{equation}\label{cd}
\vartheta = \vartheta^i_i \,.
\end{equation}

The $i-j$ component of the Einstein equations (\ref{um}) gives the evolution equation \cite{Bruni}
\begin{equation}\label{ceu}
{\vartheta^i_j}'+2\mathcal H \vartheta^i_j+\vartheta\vartheta^i_j+\frac{1}{4}(\vartheta^l_m\vartheta^m_l-\vartheta^2)\delta^i_j+\mathcal {R}^i_j-\frac{1}{4}\mathcal R\delta^i_j=0,
\end{equation}
where the Ricci tensor on the spatial hypersurfaces is given by ${}^{(3)}{R^i_j}=\mathcal {R}^i_j/a^2$ and the Ricci scalar ${}^{(3)}R=\mathcal {R}/a^2$.

The $0-0$ component of the Einstein equations gives the perturbed energy constraint
\begin{equation}\label{sei}
\vartheta^2-\vartheta^i_j\vartheta^j_i+4\mathcal H\vartheta+\mathcal R=2a^2\bar\rho_{\rm dm}\delta_{\rm dm} \,,
\end{equation}
where we define the matter density contrast
\begin{equation}
\label{deltadm}
\delta_{\rm dm}(\eta,\vec x)=\frac{\rho_{\rm dm}(\eta,\vec x)-\bar\rho_{\rm dm}(\eta)}{\bar\rho_{\rm dm}(\eta)} \,.
\end{equation}
%and an overbar denotes homogeneous background discussed below.
%
Using the $0-j$ component of the Einstein equations we find the momentum constraint
\begin{equation}\label{momentum}
{\vartheta^i_j}_{;i}=\vartheta_{,j},
\end{equation}
where a semi-colon denotes the covariant derivative with respect to the 3-metric $\gamma_{ij}$.

The perturbed Raychaudhuri equation for the expansion is found taking the trace of the evolution equation (\ref{ceu})
\begin{equation}\label{es}
\vartheta'+\mathcal H\vartheta+\vartheta^i_j\vartheta^j_i+\frac{1}{2}a^2\bar\rho_{\rm dm}\delta_{\rm dm}=0.
\end{equation}

Finally, projecting the equations $(\ref{asd})$ and $(\ref{asdw})$ parallel to $u_{\mu}$ for matter without pressure and vacuum, we obtain the energy continuity equations 
\begin{equation}\label{de}
\rho'_V=aQ,
\end{equation}
\begin{equation}\label{da}
\rho'_{\rm dm}+(3\mathcal H+\vartheta)\rho_{\rm dm}=-aQ \,.
\end{equation}
% where a prime denotes the conformal time derivative and $\mathcal H=aH$. 
%
Note that since the vacuum energy is homogeneous on comoving-orthogonal hypersurfaces we have $\rho_V=\bar\rho_V(\eta)$ and thus $Q=\bar{Q}(\eta)$. This does not imply that the vacuum energy is unperturbed but rather that we have picked a coordinate frame in which constant time hypersurfaces coincide with uniform-vacuum hypersurfaces.
In terms of the density contrast (\ref{deltadm}), the continuity equation $(\ref{da})$ becomes
\begin{equation}\label{con}
\delta_{\rm dm}'-\frac{aQ}{\rho_{\rm dm}}\delta_{\rm dm}+(1+\delta_{\rm dm})\vartheta=0.
\end{equation}

%Using de relation between Ricci scalar and three Ricci scalar $R_{(3)}$ given by
%\begin{equation}\label{po}
%a^2R=a^2R_{(3)}-2R_{00}+6\mathcal{H}^2+4\mathcal H \vartheta +\vartheta^2-\vartheta^i_j\vartheta^j_i,
%\end{equation}

\section{Background solutions}

We briefly review the solutions for the homogeneous background cosmology (\ref{aore}) with different interaction models.

%The energy conservation equation (\ref{de}) for the vacuum yields $\bar\rho_V' = a\bar Q$ and hence Eq.~$(\ref{da})$ for the dark matter density gives
%\begin{equation}
%\bar\rho_{\rm dm}' +3\mathcal{H} \bar\rho_{\rm dm} = - \bar\rho_V' \,.
%\end{equation}

The background Raychaudhuri equation is
\begin{equation}\label{bo}
\mathcal{H}'=\frac{1}{2}(2-3\Omega_{\rm dm})\mathcal{H}^2,
\end{equation}
with the dimensionless density parameter defined by $\Omega_{\rm dm}(a)=a^2\bar\rho_{\rm dm}/3\mathcal{H}^2$. The time dependence of the matter density parameter is given by
\begin{equation}\label{eq1}
\Omega_{\rm dm}'=[-3(1-\Omega_{\rm dm})+g]\mathcal H\Omega_{\rm dm},
\end{equation}
where we defined the dimensionless interaction parameter
\begin{equation}\label{inter}
g\equiv 
%- \frac{a^3Q}{3\Omega_{\rm dm}\mathcal H ^3} = 
-\frac{aQ}{\mathcal H\bar\rho_{\rm dm}} \,.
\end{equation}

%\subsection{Model with $Q=0$: $\Lambda$CDM}

For $Q=0$ there is no interaction between matter and the vacuum
%Consequently, due to equation $(\ref{de})$ 
and the vacuum energy density
%, given by
%\begin{equation}\label{L}
%\bar\rho_V=\Lambda,
%\end{equation}
is a constant in time and space, equivalent to a cosmological constant. The equation $(\ref{da})$ (with $\vartheta=0$ in the background) can be integrated to give
\begin{equation}\label{ia}
\bar\rho_{\rm dm}(a)=\rho_{\rm dm0}a^{-3},
\end{equation}
where the subscript $0$ refers to the present value, and $a_0=1$. 
%The model with a cosmological constant and matter that scales as in Eq. $(\ref{ia})$ is called 
This is the $\Lambda$CDM model.
The matter density parameter and the Hubble parameter are, respectively, given by
\begin{equation}
\Omega_{\rm dm}(a)=\frac{\Omega_{\rm dm0}}{\Omega_{\rm dm0}+(1-\Omega_{\rm dm0})a^{3}},
\end{equation}
\begin{equation}
\mathcal H (a)=a H_0\bigg[1-\Omega_{\rm dm0}+\frac{\Omega_{\rm dm0}}{a^{3}}\bigg]^{1/2},
\end{equation}
where the density parameters obey the relation $\Omega_{\rm dm}+\Omega_{V}=1$. For high-redshift (early times), as $a\ll 1$, we have a matter-dominated epoch with $\Omega_{\rm dm}\approx 1$. In the limit of large times a de Sitter vacuum dominated epoch is obtained.

More generally, the cosmological evolution for $\Omega_{\rm dm}$ and $\mathcal H$ depends of the form of the interaction parameter. In the following, we consider three different models for the possible forms of $Q$.

\subsection*{i. Model with $Q=3\alpha H\bar\rho_{\rm dm}\bar\rho_{V}/\bar\rho$}

This type of interaction corresponds to the decomposed generalized Chaplygin gas model~\cite{bento,Wands,degeneracy} where $\alpha$ is a constant parameter. The dimensionless interaction parameter $(\ref{inter})$ in this case is 
\begin{equation}
 g=-3\alpha(1-\Omega_{\rm dm}).
\end{equation}
The matter density parameter and the Hubble parameter, given by
\begin{equation}
\Omega_{\rm dm}(a)=\frac{\Omega_{\rm dm0}}{\Omega_{\rm dm0}+(1-\Omega_{\rm dm0})a^{3(1+\alpha)}},
\end{equation}
\begin{equation}
\label{HgCg}
\mathcal H (a)=a H_0\bigg[1-\Omega_{\rm dm0}+\frac{\Omega_{\rm dm0}}{a^{3(1+\alpha)}}\bigg]^{\frac{1}{2(1+\alpha)}},
\end{equation}
are solutions of the equations $(\ref{eq1})$ and $(\ref{bo})$. The standard matter era is recovered for early times ($a\ll 1$) with $\Omega_{\rm dm} \approx 1$ and $g\approx 0$.
The $\Lambda$CDM model corresponds to taking $\alpha =0$ in the above expressions. 

In the special case $\alpha=-1/2$ we have from~(\ref{HgCg}) the Hubble rate
\begin{equation}
H (a)=H_0\bigg[1-\Omega_{\rm dm0}+\frac{\Omega_{\rm dm0}}{a^{3/2}} \bigg] \,,
\end{equation}
and thus
\begin{equation}
\frac{H'}{H} = -\frac32 \mathcal H \Omega_{\rm dm} \,.
% = \frac{\rho_V'}{\rho_V}
\end{equation}
Comparing with Eq.~(\ref{de}) we see that $\bar\rho_V'/\bar\rho_V={H'}/{H}$ and thus the vacuum density decays linearly with the Hubble rate, $\bar\rho_V=2\Gamma H$, and matter is produced at a constant rate, $\dot{\bar\rho}_{\rm dm}+3H\bar\rho_{\rm dm}=\Gamma\bar\rho_{\rm dm}$~\cite{Borges, saulo}.

\subsection*{ii. Model with $Q=qH\rho_V$}

In this case the dimensionless interaction parameter $(\ref{inter})$ is\footnote{Note that $q$ here has the opposite sign to $q_V$ in Salvatelli et al \cite{Salvatelli}.}
\begin{equation}\label{comp}
g=-\left( \frac{1-\Omega_{\rm dm}}{\Omega_{\rm dm}} \right) q\,.
\end{equation}
For constant $q$ the energy continuity equation~(\ref{de}) gives
\begin{equation}\label{kai}
\bar\rho_V(a)=3H_0^2\Omega_{V 0}a^q.
\end{equation}
Substituting $(\ref{kai})$ into the Raychaudhuri equation $(\ref{bo})$ and integrating, we obtain the solution
\begin{equation}
\mathcal H(a)=H_0
%\sqrt{\frac{3\Omega_{V 0}a^{3+q}+3(1-\Omega_{V 0})+q}{(3+q)a^3}}
\sqrt{\frac{3(1-\Omega_{\rm dm0})a^{3+q}+3\Omega_{\rm dm0}+q}{(3+q)a}}
\,.
\end{equation}
The matter density parameter, given by
\begin{equation}
\Omega_{\rm dm}(a)=\frac{3\Omega_{\rm dm0}+q-q(1-\Omega_{\rm dm})a^{3+q}}{3\Omega_{\rm dm0}+q+3(1-\Omega_{\rm dm0})a^{3+q}},
\end{equation}
is solution of Eq. $(\ref{eq1})$. 
The standard matter-dominated era (Einstein-de Sitter cosmology) is recovered for early times ($a\ll 1$) with $\Omega_{\rm dm} \approx 1$ and $g\approx 0$. Note that the matter density parameter becomes negative for values $q>0$ at large times ($a\gg 1$).

The $\Lambda$CDM model corresponds to the case $q=0$.

\subsection*{iii. Model with $Q=\epsilon H\bar\rho_{\rm dm}$}

In this model the deviation from the standard evolution is given by a small constant $\epsilon$ that characterises the strength of interaction. The dimensionless interaction parameter $(\ref{inter})$ is 
\begin{equation} \label{frc}
g=-\epsilon,
\end{equation}
and for constant $\epsilon$ the equation $(\ref{da})$ (with $\vartheta=0$) can be integrated to give
\begin{equation}\label{adna}
\bar\rho_{\rm dm}(a)=\rho_{\rm dm0}a^{-(3+\epsilon)}.
\end{equation} 
Note that the matter energy density never evolves as $\bar\rho_{\rm dm}(a)\propto a^{-3}$ except for the case $\epsilon=0$, and consequently this model never has a conventional matter-dominated era. 
The amount of the vacuum energy at early times depends on the strength of interaction. 
Substituting Eq. $(\ref{adna})$ into $(\ref{de})$ gives the evolution for the vacuum energy density
\begin{equation}\label{pen}
\bar\rho_V(a)=\Lambda-\frac{\epsilon}{3+\epsilon}\bar\rho_{\rm dm}(a).
\end{equation}
Here $\Lambda$ is a constant, and the vacuum energy approaches a cosmological constant, $\bar\rho_V\to\Lambda$, as $a \rightarrow \infty$ for $\epsilon>-3$. At early times the vacuum density becomes negative for $\epsilon>0$. The $\Lambda$CDM model is recovered with zero coupling, $\epsilon=0$.

From the Friedmann equation $(\ref{aore})$ we obtain
\begin{equation}
\mathcal H (a)=a\sqrt{\frac{\rho_{\rm dm0}}{3+\epsilon}a^{-(3+\epsilon)}+\frac{\Lambda}{3}}.
\end{equation}
The dark matter density parameter is then \cite{EspanaBonet:2003vk}
\begin{equation}
\label{modeliiiOmegadm}
\Omega_{\rm dm}(a)=\frac{(3+\epsilon)\Omega_{\rm dm0}a^{-(3+\epsilon)}}{(3+\epsilon)+3\Omega_{\rm dm0}(a^{-(3+\epsilon)}-1)},
\end{equation}
At high-redshift, $a\ll 1$ for $\epsilon>-3$, the density parameter is given by
\begin{equation}\label{non}
\Omega_{\rm dm}\approx 1+\frac{\epsilon}{3}.
\end{equation}

%[NOT SURE THAT THIS IS CONSISTENT WITH WHAT YOU WROTE BEFORE:]
%If we define a ``standard expression'' for $\Omega_{\rm dm}$ as
%\begin{equation}\label{st}
%\Omega_{\rm dm}^{(s)}(a)=\frac{\Omega_{\rm dm0}^{(s)}}{\Omega_{\rm dm0}^{(s)}+[1-\Omega_{\rm dm0}^{(s)}]a^{3+\epsilon}},
%\end{equation}
%we can then write 
%\begin{equation}
%\Omega_{\rm dm}(a)=(1+\epsilon/3)\Omega_{\rm dm}^{(s)}(a) \,.
%\end{equation}

%Then we label the cases of interaction as follows: (I) correspond to the model with interaction $g=-3\alpha(1-\Omega_{\rm dm})$, (II) with $g=-q(1-\Omega_{\rm dm})/\Omega_{\rm dm}$ and (III) with $g=-\epsilon$.

\section{Growth of structure}

The metric and comoving matter density contrast can be expanded up to second order using only scalar quantities as
\begin{equation}
 \label{3metric}
\gamma_{ij}\approx[1-2\psi^{(1)}-2\psi^{(2)}]\delta_{ij}+\partial_i\partial_j\chi^{(1)}-\frac{1}{3}\nabla^2\chi^{(1)}+\partial_i\partial_j\chi^{(2)}-\frac{1}{3}\nabla^2\chi^{(2)},
\end{equation}
\begin{equation}
\delta_{\rm dm}\approx\delta_{\rm dm}^{(1)}+\frac{1}{2}\delta_{\rm dm}^{(2)}.
\end{equation}
If we assume that there are no primordial vector and tensor perturbations then the vector and tensor modes can be set to zero at first order. Vector and tensor metric perturbations will then be generated at second and higher order, but they do not affect the matter density at first or second order which is the focus of our work.

\subsection{First-order solutions}

The first order expansion of the Ricci tensor of the spatial metric (\ref{3metric}) is given by
\begin{equation}\label{tensor}
{\mathcal{R}^{(1)}}^i_j= \left( \partial^i\partial_j+\delta^i_j\nabla^2 \right) \mathcal{R}_c \,,
\end{equation}
where
\begin{equation}\label{erre}
\mathcal{R}_c=\psi^{(1)}+\frac{1}{6}\nabla^2\chi^{(1)},
\end{equation}
and thus the 3-Ricci scalar is
\begin{equation}\label{sabe'}
\mathcal{R}^{(1)}=4\nabla^2\mathcal{R}_c.
\end{equation}

The expressions $(\ref{ab})$ and $(\ref{cd})$ for the deformation tensor and scalar expansion are given to first order by
\begin{equation}\label{can}
{\vartheta^i_j}^{(1)}=-\psi^{(1)\prime}\delta^i_j+\frac{1}{2}\bigg(\partial^i\partial_j- \frac13 \delta^i_j \nabla^2 \bigg)\chi'^{(1)},
\end{equation}
\begin{equation}\label{eis}
\vartheta^{(1)}=-3\psi'^{(1)}.
\end{equation}
The momentum constraint $(\ref{momentum})$ at the first order requires
\begin{equation}
 \label{constantR}
\mathcal{R}_c'=0 \,.
\end{equation}
So $\mathcal {R}_c$ is constant in time, to be determined by initial conditions.

The continuity equation $(\ref{con})$ and Raychaudhuri equation $(\ref{es})$ for the density contrast and perturbed expansion are written up to first order as 
\begin{equation}\label{o}
\delta_{\rm dm}'^{(1)}+g\mathcal H\delta_{\rm dm}^{(1)}+{\vartheta}^{(1)}=0.
\end{equation}
\begin{equation}
 \label{thetaprime}
\vartheta'^{(1)}+\mathcal H\vartheta^{(1)}+\frac{1}{2}a^2\bar\rho_{\rm dm}\delta_{\rm dm}^{(1)}=0.
\end{equation}
subject to the first-order energy constraint $(\ref{sei})$
\begin{equation}\label{us}
4\mathcal H\vartheta^{(1)} - 2a^2\bar\rho_{\rm dm}\delta_{\rm dm}^{(1)} + \mathcal{R}^{(1)} =0,
\end{equation}

\begin{figure}
\centerline{\includegraphics[height=5.0cm]{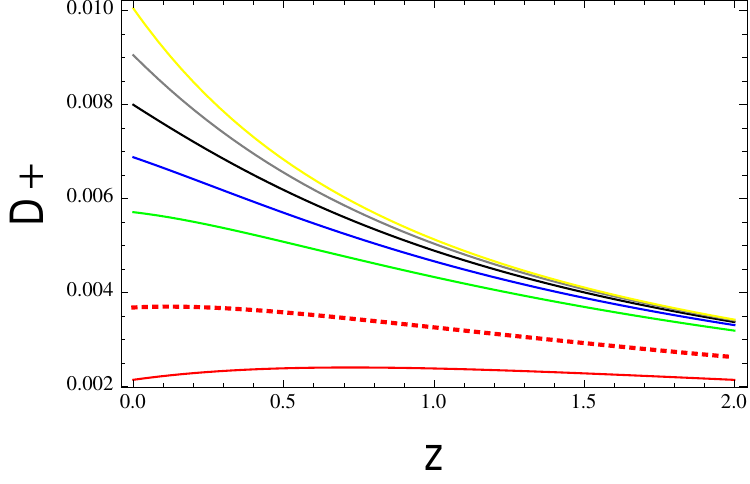} \hspace{.2in} \includegraphics[height=5.0cm]{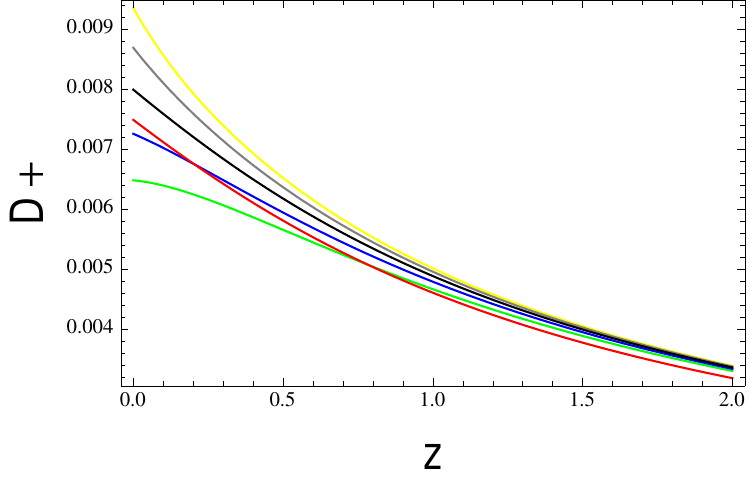}}
\caption{The first-order growing mode, $D_+$, as a function of redshift $z$ for fixed initial amplitude, $D_{+i}$. {\bf Left panel}: For model (i): $\alpha=0.2$ (yellow curve, top), $\alpha=0.1$ (grey curve), $\alpha=0$ ($\Lambda$CDM, black curve), $\alpha=-0.1$ (blue curve), and $\alpha=-0.2$ (green curve), where we have used $\Omega_{\rm dm0}=0.3$. For $\alpha=-0.5$ we used $\Omega_{\rm dm}=0.45$ (dotted red curve) and $\Omega_{\rm dm0}=0.3$ (solid red curve, bottom). {\bf Right panel}:  For model (ii): $q=0.2$ (yellow curve, top),  $q=0.1$ (grey curve), $q=0$ ($\Lambda$CDM, black curve), $q=-0.1$ (blue curve) and $q=-0.2$ (green, bottom curve at z=0), with $\Omega_{\rm dm0}=0.3$. For model (iii) we have used $\epsilon=-0.01$ (red, bottom curve for $z>1$).}
\label{fig1}
\end{figure}

Differentiating the continuity equation (\ref{o}) with respect to time and eliminating $\vartheta^{(1)}$ and $\vartheta'^{(1)}$ using the energy constraint (\ref{us}) and Raychaudhuri equation (\ref{thetaprime}), we obtain the evolution equation for the density contrast
\begin{equation}\label{four}
\delta_{\rm dm}''^{(1)}+(1+g)\mathcal H\delta_{\rm dm}'^{(1)}+\bigg[(g\mathcal H)'+g\mathcal {H}^2-\frac{1}{2}a^2\bar\rho_{\rm dm}\bigg]\delta_{\rm dm}^{(1)}=0.
\end{equation}
On the other hand, combining the first-order continuity equation $(\ref{o})$ with the constraint $(\ref{us})$, we find a first integral
\begin{equation}\label{vix}
2\mathcal H\delta_{\rm dm}'^{(1)}+\bigg[a^2\bar\rho_{\rm dm}+2g\mathcal {H}^2\bigg]\delta_{\rm dm}^{(1)} =2\nabla^2\mathcal{R}_c \,.
\end{equation}
where we used equation (\ref{sabe'}) for the first-order Ricci scalar, and we know from the momentum constraint (\ref{constantR}) that $\mathcal {R}_c$ is a constant.

The general solution for density contrast is a linear combination of growing and decaying modes. The decaying mode is the homogeneous solution to the first integral (\ref{vix}), i.e., setting the $\mathcal{R}_c$ to zero. Neglecting this decaying mode, we are left with the growing mode driven by the non-zero Ricci curvature
\begin{equation}\label{ef}
\delta_{\rm dm}^{(1)}(\eta,\vec x)=C(\vec x)D_+(\eta).
\end{equation} 
where we have from $(\ref{vix})$
\begin{equation}\label{sao1}
C(\vec x) = \bigg( f_{1i} + \frac32\Omega_{\rm dm,i} + g_i \bigg)^{-1} \frac{\nabla^2\mathcal{R}_c}{\mathcal{H}_i^2 D_{+i}} \,,
\end{equation}
and we define the linear growth rate as
\begin{equation}\label{gr}
f_1=\frac{D_+'}{\mathcal H D_+} \,.
\end{equation} 
The growing mode is then
\begin{equation}\label{sao}
D_+(\eta)= \bigg(\frac{\mathcal{H}_i}{\mathcal H}\bigg)^2 \bigg( f_{1i} + \frac32 \Omega_{\rm dm,i} +g_i \bigg) \bigg(f_1+\frac{3\Omega_{\rm dm}}{2}+g\bigg)^{-1}  D_{+i} \,.
\end{equation}
Note that in this expression for the growing mode we have left an arbitrary overall normalisation constant, $D_{+i}$.

If we set initial conditions at high redshift, $a_i\ll1$, during a standard matter-dominated era, where $\Omega_{dm i} = 1$, $f_{i}=1$ and $g_{i}=0$, then we have 
\begin{equation}
C(\vec x)=\frac{2}{5}\frac{\nabla^2\mathcal{R}_c}{\mathcal{H}_i^2 D_{+i}}  \,,
\end{equation}
and the growing mode (\ref{sao}) reduces to
\begin{equation}
D_+(\eta) = \frac{5}{2}\bigg(\frac{\mathcal{H}_i}{\mathcal H}\bigg)^2 \bigg(f_1+\frac{3\Omega_{\rm dm}}{2}+g\bigg)^{-1} D_{+i} \,.
\end{equation}
From $(\ref{ef})$ the first-order solution is then
\begin{equation}\label{pa} 
\delta_{\rm dm}^{(1)}(\eta,\vec x)= \bigg(f_1+\frac{3\Omega_{\rm dm}}{2}+g\bigg)^{-1} \frac{\nabla^2\mathcal{R}_c}{\mathcal{H}^2} \,.
\end{equation}

Substituting the growing mode solution (\ref{ef}) and (\ref{gr}) in the continuity equation $(\ref{o})$ we obtain the expansion scalar
\begin{equation}\label{mar}
\vartheta^{(1)} = - (f_1+g) \mathcal H \delta_{\rm dm}^{(1)} \,.
\end{equation}
The metric perturbation $\psi^{(1)}$ is given by integrating Eq.~(\ref{eis}). Using (\ref{o}) and (\ref{pa}) we obtain
\begin{equation}\label{mn}
\psi^{(1)}=\mathcal{R}_c+\frac{1}{3}\nabla^2\mathcal{R}_c\bigg[\frac{1}{\mathcal{H}^2}\bigg(f_1+\frac{3}{2}\Omega_{\rm dm}+g\bigg)^{-1}+\int\frac{g}{\mathcal{H}}\bigg(f_1+\frac{3}{2}\Omega_{\rm dm}+g\bigg)^{-1}d\eta\bigg].
\end{equation}
Equation $(\ref{erre})$ then gives
\begin{equation}\label{eisa}
\chi^{(1)}=-2\mathcal{R}_c\bigg[\frac{1}{\mathcal{H}^2}\bigg(f_1+\frac{3}{2}\Omega_{\rm dm}+g\bigg)^{-1}+\int\frac{g}{\mathcal{H}}\bigg(f_1+\frac{3}{2}\Omega_{\rm dm}+g\bigg)^{-1}d\eta\bigg].
\end{equation}
For completeness we note that the expression for the deformation tensor ${\vartheta^i_j}^{(1)}$ is then given by $(\ref{can})$. 

The expressions above are valid only if the matter flow follows geodesics, as we have assumed throughout. 
For a dimensionless parameter interaction $g$ equal to zero the results for the $\Lambda$CDM model are recovered~\cite{Bruni}.

Figure~\ref{fig1} shows the plot of the evolution of first-order growing mode $D_+$ for the  $\Lambda$CDM and all three interaction models obtained by solving the differential equation $(\ref{four})$ with the same initial amplitude $D_{+i}$ for all of the growing modes at $z=1000$. 
%In the left panel we show the model (i) that corresponds to a decomposed generalized Chaplygin gas. Models (ii) and (iii) are shown in the right panel. 
%
When $g>0$ we have energy flux from vacuum to dark matter, since $Q<0$, and dark matter is created. In this case the first-order growing mode is suppressed with respect to the $\Lambda$CDM model (black curve) for a given value of the present day dark matter, $\Omega_{\rm dm0}$. 
This is because the dark matter density is lower at early times when we fix the dark matter density today.
When $g<0$ we have energy flow from dark matter to the vacuum, since $Q>0$, and dark matter is annihilated or decays. In this case there is an enhancement in the first-order growing mode for the same value of $\Omega_{\rm dm0}$ \cite{Wang:2013qy}. 

%[WE MIGHT TRY SHOWING THE GROWING MODE FOR MODELS WITH A FIXED VALUE OF $\Omega_{\rm dm}$ AT $z=1000$ WHERE WE MIGHT SEE THE OPPOSITE EFFECT: THE GROWING MODE IS THEN ENHANCED AT LATE TIMES FOR POSITIVE $g$.]

\begin{figure*}
\centerline{\includegraphics[height=5.0cm]{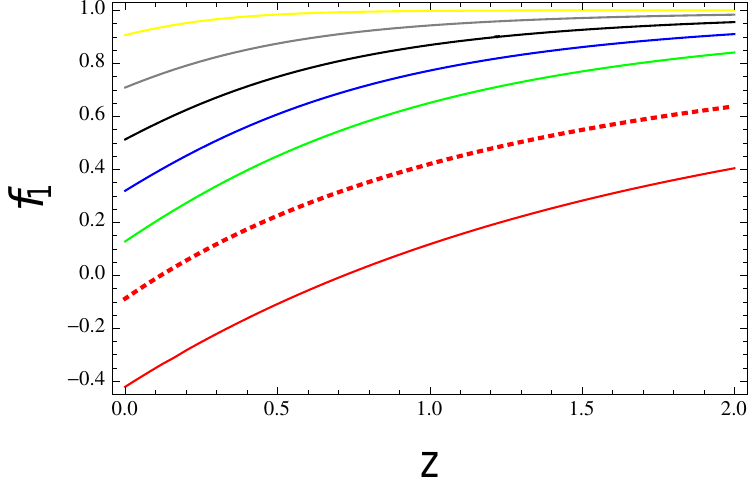} \hspace{.2in} \includegraphics[height=5.0cm]{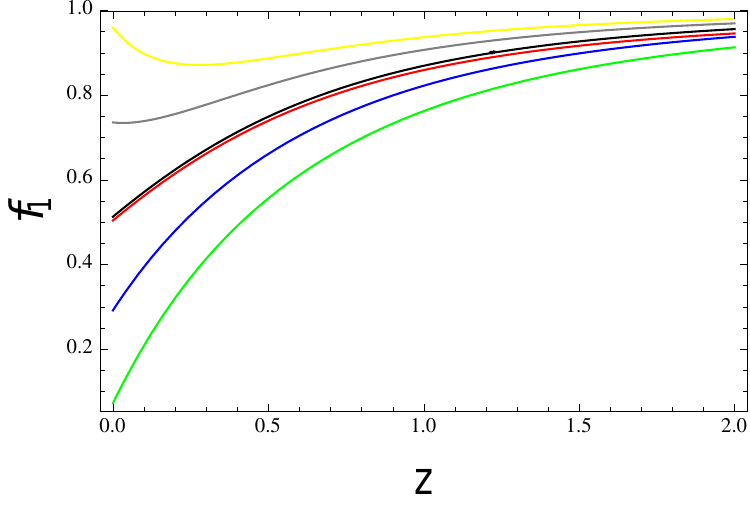}}
\caption{{\bf Left panel}: The first-order growth rate, $f_1$ defined in Eq.~(\ref{gr}), for $\Lambda$CDM model (black curve) and for model (i): $\alpha=0.2$ (yellow curve, top), $\alpha=0.1$ (grey curve), $\alpha=0$ ($\Lambda$CDM, black curve), $\alpha=-0.1$ (blue curve) and $\alpha=-0.2$ (green curve), where we have used $\Omega_{\rm dm0}=0.3$. For $\alpha=-0.5$ we used $\Omega_{\rm dm}=0.45$ (dotted red curve) and $\Omega_{\rm dm0}=0.3$ (solid red curve, bottom). 
{\bf Right panel}: model (ii): $q=0.2$ (yellow curve, top),  $q=0.1$ (grey curve), $q=0$ ($\Lambda$CDM, black curve), $q=-0.1$ (blue curve) and $q=-0.2$ (green curve, bottom). For model (iii) we have used $\epsilon=-0.01$ with $\Omega_{\rm dm0}=0.3$ (red curve).}
\label{fig2}
\end{figure*}

In figure~\ref{fig2} we plot the evolution of the growth rate $f_1$ defined in Eq.~(\ref{gr}) for model (i) (left panel) and for models (ii) and (iii) (right panel) with different values for the model parameters $\alpha$, $q$ and $\epsilon$.

\subsection{Redshift-space distortions}

\begin{figure*}
\centerline{\includegraphics[height=5.0cm]{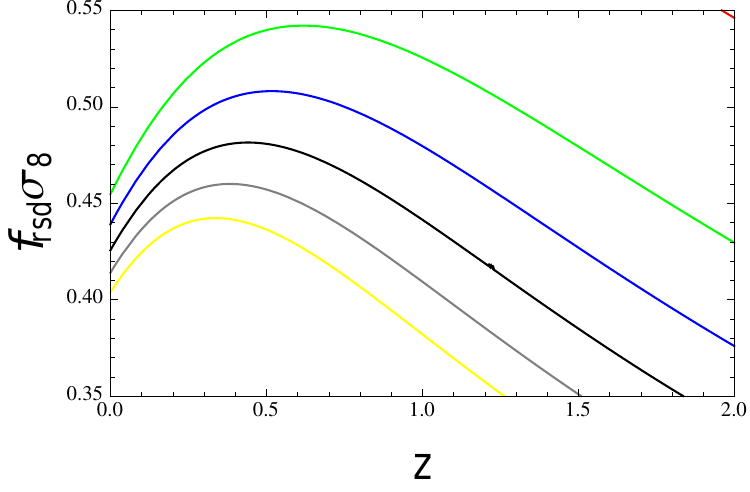} \hspace{.2in} \includegraphics[height=5.0cm]{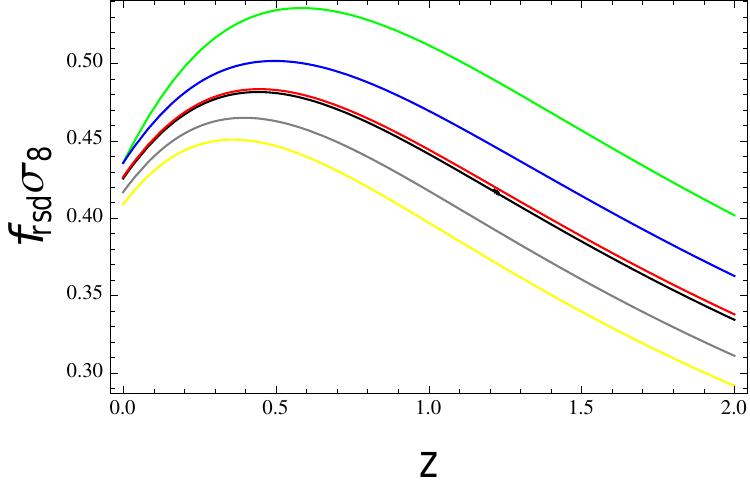}}
\caption{Magnitude of redshift space distortions for dark matter, $f_{\rm rsd}\sigma_{8}$ given in (\ref{frsd}), versus redshift, $z$, normalised to $\sigma_{8}=0.83$ at present. {\bf Left panel}: model (i): $\alpha=-0.2$ (green curve, top), $\alpha=-0.1$ (blue curve), $\alpha=0$ ($\Lambda$CDM, black curve), $\alpha=0.1$ (grey curve) and $\alpha=0.2$ (yellow curve, bottom) all with $\Omega_{\rm dm0}=0.3$.
 {\bf Right panel}: model (ii): $q=-0.2$ (green curve, top), $q=-0.1$ (blue curve), $q=0$ ($\Lambda$CDM, black curve), $q=0.1$ (grey curve) and $q=0.2$ (yellow curve, bottom). For model (iii) we have used $\epsilon=-0.01$ with $\Omega_{\rm dm0}=0.3$ (red curve).}
\label{fig3}
\end{figure*}

Redshift-space distortions (RSD) arise from peculiar velocities of galaxies, i.e., the perturbed expansion, $\vartheta$, given in (\ref{cd}). This induces an anisotropy in the apparent clustering of galaxies in redshift space, where we use the observed redshift to determine the radial distance. This observed anisotropy thus provides information about the formation of large-scale structure \cite{Kaiser:1987qv}. 

In standard $\Lambda$CDM (where the dimensionless parameter $g=0$) 
the variance of the expansion is usually characterised from equation ($\ref{mar}$) by~\cite{three}
%the bias-independent combination 
\begin{equation}
\langle \vartheta^2/{\cal H}^2 \rangle^{1/2} = f_1(z)\sigma_{8}(z) \,,
\end{equation}
where $f_1(z)$ is the linear growth rate and $\sigma_{8}(z)=\langle \delta_m^2 \rangle^{1/2}$ is the rms mass fluctuation in a sphere with comoving radius $8h^{-1}$Mpc, used to describe the amplitude of density perturbations. If we use the growing mode normalised to unity today, $D^{N}_{+}(z)=\delta_{\rm dm}(z)/\delta_{\rm dm}(0)$, then we can write $\sigma_{8}(z)=\sigma_8(0)D^{N}_{+}(z)$ where $\sigma_8(0)$ gives the present rms matter fluctuations. 

More generally, for interacting models, the dimensionless interaction parameter $g$ contributes explicitly in equation ($\ref{mar}$) for redshift space distortions. If we assume that galaxies still trace the motion of the underlying dark matter (i.e., neglecting any velocity bias) then the variance of the expansion ($\ref{mar}$) is given by
\begin{equation}
\label{deffrsd}
\langle \vartheta^2/{\cal H}^2 \rangle^{1/2} = f_{\rm rsd}(z)\sigma_{8}(z)
%=[f_1+g](z)\sigma_{8}(0)D^{N}_{+}(z)
\,,
\end{equation}
where
\begin{equation}\label{frsd}
f_{\rm rsd}(z)
%\sigma_{8}(z) 
= f_1(z)+g(z)
% [f_1(z)+g(z)] \sigma_{8}(0)D^{N}_{+}(z)
\,.
\end{equation}

Figure~\ref{fig3} shows the theoretical predictions for $f_{\rm rsd}\sigma_8$ as a function of redshift $z$ for the different interacting models, where we fix $\sigma_{8}(0)=0.83$ \cite{planck}. 
We see that in contrast to the linear growth rate, the RSD distortions are enhanced by energy transfer from the vacuum to dark matter. The peculiar velocity field responds to the local gravitational potential and thus the total comoving density perturbation, not just the density contrast. 

\begin{figure}
\centerline{\includegraphics[height=5.0cm]{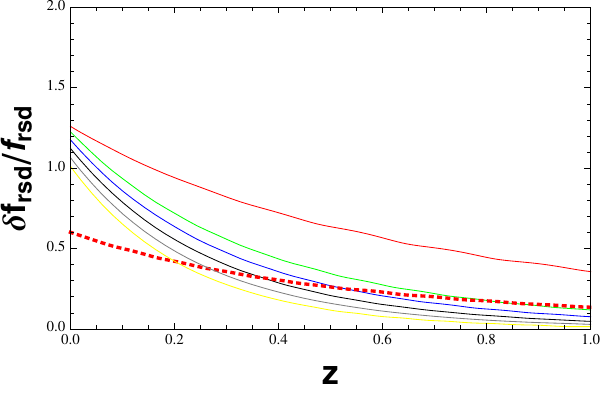} \hspace{.2in} \includegraphics[height=5.0cm]{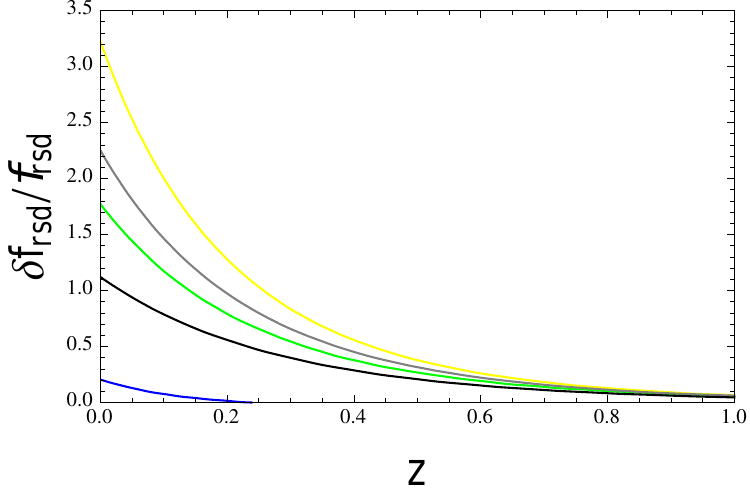}}
\caption{{\bf Left panel}: Plot of the relative percentage difference 
%$\left(\frac{\Omega_{\rm dm}^{ {\gamma_{\rm rsd}}}}{f_1+g}-1\right)\times 100$ 
between the analytical approximation (\ref{defalpha}) for $f_{\rm rsd}$ and the numerical solution 
%$f_1+g$, 
in model (i) for $\alpha=-0.2$ (green), $\alpha=-0.1$ (blue), $\alpha=0$ (black), $\alpha=0.1$ (grey) and $\alpha =0.2$ (yellow) with $\Omega_{\rm dm}=0.3$, 
$\alpha=- 0, 5$ (dotted red curve) and $\alpha=-0.5$ (solid red curve), with $\Omega_{\rm dm}= 0.45$. 
{\bf Right panel}: For model (ii) where $q =0.2$ (yellow, top), $q=0.1$ (grey), $q=-0.2$ (green curve), $q=0$ ($\Lambda CDM$, black) and $q=-0.1$ (blue curve, bottom). %the approximation is worse for positive q.
For the model (iii) we have plotted for $\epsilon=-0.01$ (red). }
\label{fig4}
\end{figure}

% \subsubsection{Linear growth rate, $f_1$}

The second-order differential equation for the density contrast ($\ref{four}$) can be written as a first-order differential equation for the redshift-space distortion parameter
%, $f_1$, 
\begin{equation}
\label{raiz-eta}
2{\mathcal H}^{-1} f_{\rm rsd}' + (2f_{\rm rsd}+4-3\Omega_{\rm dm}-2g)f_{\rm rsd} = 3\Omega_{\rm dm} \,.
\end{equation}
In the conventional matter-dominated era at high redshift with $\Omega_{\rm dm}=1$ and the dimensionless parameter interaction $g=0$, we have a solution corresponding to the standard growing mode\footnote{Note we also have a solution $f_{\rm rsd}=f_1=-3/2$ corresponding to the standard decaying mode.} with $f_{\rm rsd}=f_1=1$ and the linear growing mode is proportional to the scale factor, $D_+\propto a$.  This describes the early growing mode at high redshifts as $g\to0$ and $\Omega_{\rm dm}\to1$ in models (i) and (ii), as well as $\Lambda$CDM.

More generally, when vacuum energy contributes to the total density ($\Omega_{\rm dm}<1$) we can express the first-order equation (\ref{raiz-eta}) for the RSD parameter as a function of the density parameter, written in terms of $\Omega_V=1-\Omega_{\rm dm}$,
\begin{equation}
\label{raiz}
%6(\Omega_{\rm dm}-1)\Omega_{\rm dm}\bigg[\frac{df_1}{d\Omega_{\rm dm}}+\frac{3}{2}\bigg]+(2f_1+3\Omega_{\rm dm})(f_1+2-3\Omega_{\rm dm})=-g\bigg[2f_1+2\Omega_{\rm dm}\frac{df_1}{d\Omega_{\rm dm}}+2\frac{g'}{g\mathcal H}-3\Omega_{\rm dm}+4\bigg],
% \left[ -6(1-\Omega_{\rm dm}) + 2g \right] \Omega_{\rm dm} \frac{d}{d\Omega_{\rm dm}} \left( f+g \right) = (2f+4-3\Omega_{\rm dm})(f+g) = 3\Omega_{\rm dm}
2\left( 3\Omega_V - g \right) (1-\Omega_V) \frac{d}{d\Omega_V} f_{\rm rsd} + (2f_{\rm rsd}+1+3\Omega_V-2g)f_{\rm rsd} = 3(1-\Omega_V)
 \,.
\end{equation}
Note that $g$ is a given function of the density parameter, $\Omega_V$, in each of our interaction models.

For $\Omega_{\rm dm}=1$ to be a fixed point of Eq.~(\ref{eq1}) we require $g=0$ when $\Omega_V=0$.
If we then expand the dimensionless interaction parameter (\ref{inter}) as a Taylor series about the standard matter-dominated ($\Omega_{\rm dm}=1$, $\Omega_V=0$) solution
\begin{equation}
g = g_1 \Omega_V 
%+ \frac12 g_2 \Omega_V^2 
+ \ldots \,,
\end{equation}
we obtain an expression for the redshift-distortion parameter ($\ref{frsd}$) 
\begin{equation}
\label{Taylorf}
f_{\rm rsd} = f_{\rm rsd,0}+ f_{\rm rsd,1} \Omega_V 
%f_1 = f_{1,0} + f_{1,1} \Omega_V 
%+ \frac12 f_{1,2} \Omega_V^2 
+ \ldots 
\,,
\end{equation}
From Eq.~(\ref{raiz}) we require
\begin{eqnarray}
 \label{simultaneous}
%(1+2f_{1,0})(f_{1,0}+g_0) - 2g_0(f_{1,1}+g_1) &=& 3 \nonumber \\
(1+2f_{\rm rsd,0})f_{\rm rsd,0} &=& 3 \nonumber \\
(3-2g_1+2f_{\rm rsd,1})f_{\rm rsd,0} + (1+2f_{\rm rsd,0})f_{\rm rsd} + 2(3-g_1)f_{\rm rsd} &=& -3 \,.
\end{eqnarray}

For $\Lambda$CDM with $g=0$ we have from (\ref{simultaneous}) 
\begin{eqnarray}
 \label{simultaneousLCDM}
(1+2f_{\rm rsd,0})f_{\rm rsd,0} &=& 3 \nonumber \\
(3+2f_{\rm rsd,1})f_{\rm rsd,0} + (1+2f_{\rm rsd,0})f_{\rm rsd,1} + 6f_{\rm rsd,1} &=& -3 \,.
\end{eqnarray}
This gives either $f_{\rm rsd,0}=-3/2$ (decaying mode) or $f_{\rm rsd,0}=1$ (growing mode) and then $f_{\rm rsd,1}=-6/11$, corresponding to \cite{Peebles}
\begin{equation}
\label{Peebles-index}
f_1 = f_{\rm rsd} \approx \Omega_{\rm dm}^{6/11} \,.
\end{equation}

More generally, we can give a similar approximation for the RSD parameter in terms of $\Omega_{\rm dm}$ when $g\neq0$. In models (i) or (ii) we write
\begin{equation}
 \label{defalpha}
f_{\rm rsd} \approx \Omega_{\rm dm}^{\gamma} \,.
\end{equation}
For model (i) we have $g=-3\alpha\Omega_V$ and hence $g_1=-3\alpha$ in Eq.~(\ref{simultaneous}). Thus we have for the growing mode $f_{\rm rsd,0}=1$ and $f_{\rm rsd,1}=-\gamma$ such that
\begin{equation}\label{gCgtilde}
\gamma
%= \frac{6-27\alpha-18\alpha^2}{11+6\alpha} + 3\alpha 
= \frac{6+6\alpha}{11+6\alpha} \,,
\end{equation}
For model (ii) we have $g=-q\Omega_V(1-\Omega_V)^{-1}$ and hence $g_1=-q$ in Eq.~(\ref{simultaneous}). Thus we have $f_{\rm rsd,0}=1$ and $f_{\rm rsd,1}=-\gamma$ where in this case 
\begin{equation}\label{m2tilde}
\gamma
%= \frac{6-9q-2q^2}{11+2q} + q 
= \frac{6+2q}{11+2q} \,. 
\end{equation}
Note that for a given value of $\Omega_{\rm dm}$ the RSD index $\gamma$, is now enhanced for $\alpha>0$ in Eq.~(\ref{gCgtilde}) and $q>0$ in (\ref{m2tilde}), corresponding to $g<0$. 
%This is in contrast to the growth rate $f_1=f_{\rm rsd}-g$ index $\gamma$ in the corresponding Eqs.~(\ref{gCggamma}) and $q>0$ in (\ref{qgamma}).

As shown in figure~\ref{fig4}, the analytical formula (\ref{defalpha}) for the RSD parameter can be used as a good approximation for model (i), corresponding to the decomposed generalized Chaplygin gas, just as it is used in $\Lambda$CDM. For this class of model the expression (\ref{defalpha}) with the growth index (\ref{gCgtilde}) works very well within an error less than 1.5 percent up to redshift $z=0$ for $|\alpha|<0.5$. On the other hand, for model (ii) shown in the right panel of figure~\ref{fig4}, the expression (\ref{defalpha}) with the growth index (\ref{m2tilde}) is a good approximation with errors below 3.5\% for $|q|<0.2$.  
In all the cases shown, the approximations for $f_{\rm rsd}$ become extremely accurate when applied for higher redshift where $1-\Omega_{\rm dm}\ll1$.

Finally, for model (iii) $g=-\epsilon$ and thus is not zero at early times so $\Omega_{\rm dm}\neq1$ at high redshift. Instead from Eq.~(\ref{modeliiiOmegadm}) we have $\Omega_{\rm dm}\to1+(\epsilon/3)$. Nonetheless, from Eq.~(\ref{raiz-eta}), we see that there is still an early time solution for the RSD parameter $f_{\rm rsd}\to f_{\rm rsd,0}=1$ as $\Omega_{\rm dm}\to1+(\epsilon/3)$\footnote{We also find a decaying mode solution at early times in this model corresponding to $f_1=-(3-\epsilon)/2$ and $f_{\rm rsd}=-(3+\epsilon)/2$}.
This corresponds to an early-time growing mode solution $D_+\propto a^{1+\epsilon}$ with modifield growth rate $f_1=1+\epsilon$.
%
%Taking the same Taylor series expansion as previously for the RSD parameter (\ref{Taylorf}) 
Expanding about this early-time solution 
%we can define
%\begin{equation}
%\Omega_{\rm dm} = \left( 1+\frac{\epsilon}{3} \right) (1 - \Omega_V) \,,
%\end{equation}
%and the growth rate is then given by 
%\begin{equation}
%f_1 = f_{1,0} + f_{1,1} \Omega_V 
%+ \ldots 
%\,.
%\end{equation}
%Rewriting Eq.~(\ref{raiz}) as
%\begin{equation}
%2(3+\epsilon) \Omega_V (1-\Omega_V) \frac{df_1}{d\Omega_V} + \left[ 2f_1+4-(3+\epsilon)(1-\Omega_V) \right] (f_1-\epsilon) = (3+\epsilon)(1-\Omega_V)
%\,,
%\end{equation}
%we find
%\begin{equation}
%f_{1,1} = - \frac{(3+\epsilon)(f_{1,0}+1-\epsilon)}{4f_{1,0}+7-\epsilon}
%\,.
%\end{equation}
%Thus we may write the linear growth rate close to the high-redshift limit, $\Omega_{\rm dm}=1+(\epsilon/3)$, as
%\begin{equation}
%\label{f1approxepsilon}
%f_1 \approx \left( \frac{\Omega_{\rm dm}}{1+(\epsilon/3)} \right)^{\gamma} + \epsilon \,,
%\end{equation}
%where the index is 
%\begin{equation}
%\label{gammaapproxepsilon}
%\gamma = \frac{6+2\epsilon}{11+3\epsilon} 
% \,.
%\end{equation}
%For $\epsilon=0$ we recover the $\Lambda$CDM result (\ref{Peebles-index}).
we find an analogous approximation for the RSD parameter (\ref{defalpha})
%, can be written as
\begin{equation}
 \label{defalphaIII}
f_{\rm rsd} \approx \left( \frac{\Omega_{\rm dm}}{1+(\epsilon/3)} \right)^\gamma \,,
\end{equation}
where the index $\gamma$ is given by
\begin{equation}
\label{gammaapproxepsilon}
\gamma = \frac{6+2\epsilon}{11+3\epsilon} 
 \,.
\end{equation}
For $\epsilon=0$ we recover the $\Lambda$CDM result (\ref{Peebles-index}).

\subsection{Second-order perturbations}

To investigate the emergence of nonlinear structure in the presence of energy transfer we consider the second-order terms in the continuity equation $(\ref{con})$ and Raychaudhuri equation $(\ref{es})$ for the evolution of the density contrast and perturbed expansion in comoving synchronous coordinates
\begin{equation}\label{two}
\delta_{\rm dm}'^{(2)}+g\mathcal H\delta_{\rm dm}^{(2)}+\vartheta^{(2)} = -2\delta_{\rm dm}^{(1)}\vartheta^{(1)} \,,
\end{equation}
\begin{equation}\label{tre}
\vartheta'^{(2)}+\mathcal{H}\vartheta^{(2)}+\frac{1}{2}a^2\bar\rho_{\rm dm}\delta_{\rm dm}^{(2)}= -2{\vartheta^{(1)}}^i_j{\vartheta^{(1)}}^j_i \,,
\end{equation}
subject to the constraint $(\ref{sei})$
\begin{equation}\label{one}
4\mathcal{H}\vartheta^{(2)}-2a^2\bar\rho_{\rm dm}\delta_{\rm dm}^{(2)}+\mathcal{R}^{(2)}=2{\vartheta^{(1)}}^i_j{\vartheta^{(1)}}^j_i-2{\vartheta^{(1)}}^2.
\end{equation}
The left-hand-sides of these equations have the same form as the first-order equations (\ref{thetaprime}), (\ref{us}) and (\ref{o}), but now with source terms on the right-hand-sides of the equations that are quadratic in the first-order quantities.

Differentiating the continuity equation ($\ref{two}$) with respect to time and eliminating $\vartheta'^{(2)}$ and $\vartheta^{(2)}$ using the Raychaudhuri equation~($\ref{tre}$) and constraint~($\ref{two}$), we obtain an evolution equation for the second-order density contrast
\begin{equation}\label{demais}
\delta_{\rm dm}''^{(2)}+(1+g)\mathcal H\delta_{\rm dm}'^{(2)}+\bigg[(g\mathcal H)'+g\mathcal {H}^2-\frac{1}{2}a^2\bar\rho_{\rm dm}\bigg]\delta_{\rm dm}^{(2)}=-2\mathcal H\delta_{\rm dm}^{(1)}\vartheta^{(1)}-2\delta_{\rm dm}'^{(1)}\vartheta^{(1)}-2\delta_{\rm dm}^{(1)}\vartheta'{(1)}+2{\vartheta^{(1)}}^i_j{\vartheta^{(1)}}^j_i.
\end{equation}
The differential equation (\ref{demais}) for the second-order density contrast has a particular solution, $\delta_{\rm dm,p}^{(2)}$, driven by the second-order source terms on the right-hand-side. However the general solution also includes the decaying and growing mode solutions to the homogeneous (source-free) equation, i.e., with the right-hand-side set to zero, with two arbitrary constants of integration. Since the source-free equation is the same as the first-order equation (\ref{four}), the homogeneous growing and decaying modes have the same time-dependence as the first-order solutions, but with second-order coefficients, to be set by the initial conditions.

As we did for the first-order equations, we can combine the constraint $(\ref{one})$ and the continuity equation $(\ref{two})$ to obtain a first integral
\begin{equation}\label{me}
4\mathcal H\delta_{\rm dm}'^{(2)}+2\bigg[a^2\rho_{\rm dm}+2g\mathcal H^2\bigg]\delta_{\rm dm}^{(2)}-{\mathcal {R}}^{(2)}=2{\vartheta^{(1)}}^2-2{\vartheta^{(1)}}^i_j{\vartheta^{(1)}}^j_i-8\mathcal{H}\delta_{\rm dm}^{(1)}\vartheta^{(1)}.
\end{equation}
Here, and in (\ref{one}), the second-order part of the comoving curvature is given by \cite{matarrese,Bruni} 
%[DW - I ADDED A FACTOR $1/2$ ON THE LHS TO MATCH EQ.(3.50) IN BRUNI ET AL \cite{Bruni}]
\begin{align}\label{biss}
\frac12
\mathcal{R}^{(2)} = 2\nabla^2\bigg[\psi^{(2)}+\frac{1}{6}\nabla^2\chi^{(2)}\bigg] + 6\partial^i\psi^{(1)}\partial_i\psi^{(1)}
+ 16\psi^{(1)}\nabla^2\psi^{(1)}+4\psi^{(1)}\partial_i\partial_j{\chi^{(1)}}^{ij}-2\partial_i\partial_j\psi^{(1)}{\chi^{(1)}}^{ij}+\nonumber \\+{\chi^{(1)}}^{ij}\nabla^2{\chi^{(1)}}_{ij}-2{\chi^{(1)}}^{jk}\partial_l\partial_k{\chi^{(1)}}^l_j-\partial_l{\chi^{(1)}}^{lk}\partial_j{\chi^{(1)}}^j_k+\frac{3}{4}\partial_k{\chi^{(1)}}^{lj}\partial^k{\chi^{(1)}}_{lj}-\frac{1}{2}\partial_k{\chi^{(1)}}^{lj}\partial_l{\chi^{(1)}}^k_j.
\end{align}
Unlike the first-order case, the second-order comoving scalar is no longer constant on all scales. However to leading order in a spatial gradient expansion we have \cite{Bruni:2014xma} 
\begin{equation}\label{biss-largescale}
\frac12
\mathcal{R}^{(2)}=2\nabla^2 \psi^{(2)}  +6\partial^i\psi^{(1)}\partial_i\psi^{(1)} + 16\psi^{(1)}\nabla^2\psi^{(1)} + {\cal O}(\nabla^4)
\,,
\end{equation}
and this does remain constant in the large-scale limit \cite{Malik:2003mv,Langlois:2005qp}.

As in the first-order case, 
%the general solution of the equation (\ref{me}) is given by a linearly decaying mode, which we neglect, 
we may neglect the decaying mode for regular initial conditions, while the amplitude of the  
%and growing part driven by the Ricci curvature and, at second order,
%the additional terms on the right-hand-side of (\ref{me}), quadratic in the first-order quantities $\vartheta^{(1)}$ and $\delta_{\rm dm}^{(1)}$. This growing part contains a 
homogeneous growing mode must be set from the constraint equation (\ref{me}).
The homogeneous, linearly-growing mode, $\delta_{\rm dm,h}^{(2)}\propto D_+$,  is driven by the constant part of the second-order curvature, $\mathcal{R}^{(2)}_{\rm h}=$constant
%. Following the approach in \cite{Bruni}, we refer to this as the homogeneous part, as it corresponds to the homogeneous solution to the second-order differential equation (\ref{demais}) for the density contrast, $\delta_{\rm dm,p}^{(2)}$. The general solution to (\ref{demais}) also contains 
while at second-order there is also the particular solution, $\delta_{\rm dm,p}^{(2)}$, 
%driven by the second-order terms on the right-hand-side. This 
corresponding to a solution to (\ref{me}) sourced by the time-dependent part of the comoving curvature, $\mathcal{R}^{(2)}_{\rm p}=\mathcal{R}^{(2)}-\mathcal{R}^{(2)}_{\rm h}$.
%, and the quadratic terms on the right-hand-side of equation~(\ref{me}).

Note that the homogeneous, linearly-growing mode, $\delta_{\rm dm,h}^{(2)}={\cal O}(\nabla^2/\mathcal{H}^2)$, will dominate on large scales where the comoving curvature perturbation (\ref{biss-largescale}) is constant. The particular, nonlinearly-growing solution, $\delta_{\rm dm,p}^{(2)}={\cal O}(\nabla^4/\mathcal{H}^4)$, will dominate on smaller scales and late times.

\subsubsection{Particular solution}

The time-dependent part of comoving Ricci scalar $\mathcal{R}^{(2)}$ can be obtained by differentiating $(\ref{one})$ with respect to time. 
After some calculation, using the equations for the second-order continuity equation (\ref{two}) and Raychaudhuri equation (\ref{tre}) as well as the Einstein evolution equation (\ref{ceu}) to first order, we obtain
\begin{equation}
\label{R2prime}
\mathcal{R}'^{(2)} = -2 \mathcal {R}^{j(1)}_i \partial^i\partial_j\chi'^{(1)}
%-2[\partial^i\partial_j\chi'^{(1)}\partial^j\partial_i\mathcal{R}_c+\nabla^2\chi'^{(1)}\nabla^2\mathcal{R}_c]
\,,
\end{equation}
where the first-order Ricci tensor on the comoving spatial hypersurfaces, $\mathcal {R}^{i(1)}_j=[\partial^i\partial_j+\delta^i_j\nabla^2]\mathcal{R}_c$, is constant in time.
Integrating (\ref{R2prime}), and using the solution $(\ref{eisa})$ for $\chi^{(1)}$, we find
\begin{equation}\label{sabe-partic}
\mathcal{R}^{(2)}_{\rm p}=4\bigg[\frac{1}{\mathcal{H}^2}\bigg(f_{\rm rsd}+\frac{3}{2}\Omega_{\rm dm}\bigg)^{-1}+\int \frac{g}{a\mathcal{H}^2}\bigg(f_{\rm rsd}+\frac{3}{2}\Omega_{\rm dm}\bigg)^{-1}da\bigg][\partial^i\partial_j\mathcal{R}_c\partial^j\partial_i\mathcal{R}_c+(\nabla^2\mathcal{R}_c)^2].
\end{equation}
Note that this time-dependent part of the Ricci scalar at second order is fourth-order in spatial derivatives, consistent with our earlier conclusion that the Ricci scalar is constant at leading order on large scales (\ref{biss-largescale}). 

The constraint equation $(\ref{me})$ for the particular solution to equation (\ref{demais}) with the time-dependent part of the Ricci scalar, $\mathcal{R}^{(2)}_{\rm p}$:
\begin{equation}
4\mathcal H\delta_{\rm dm,p}'^{(2)}+2\bigg[a^2\bar\rho_{\rm dm}+2g\mathcal H^2\bigg]\delta_{\rm dm,p}^{(2)}
= {\mathcal {R}}^{(2)}_{\rm p} + 2{\vartheta^{(1)}}^2-2{\vartheta^{(1)}}^i_j{\vartheta^{(1)}}^j_i-8\mathcal{H}\delta_{\rm dm}^{(1)}\vartheta^{(1)} \,,
\end{equation}
can thus be written as
\begin{equation}
\label{deu}
4\mathcal H\delta_{\rm dm,p}'^{(2)}+2\bigg[a^2\bar\rho_{\rm dm}+2g\mathcal H^2\bigg]\delta_{\rm dm,p}^{(2)}
=  \mathcal{S}(a, \Sigma) \frac{(\nabla^2\mathcal{R}_c)^2}{\mathcal H^2} \,,
\end{equation}
where we introduce the dimensionless shape coefficient
\begin{equation}
\Sigma(\vec x)=\frac{\vartheta^i_j\vartheta^j_i}{\vartheta^2}=\frac{\partial^i\partial_j\mathcal{R}_c\partial^j\partial_i\mathcal{R}_c}{(\nabla^2\mathcal{R}_c)^2},
\end{equation}
and define the dimensionless source function
\begin{eqnarray}\label{func}
\mathcal{S}(a,\Sigma) &=& \frac{2f_{\rm rsd}^2(1-\Sigma)+8f_{\rm rsd}+4(f_{\rm rsd}+\frac{3}{2}\Omega_{\rm dm})(1+\Sigma)}{(f_{\rm rsd}+\frac{3}{2}\Omega_{\rm dm})^2}\nonumber \\
&&
+ 4(1+\Sigma) \mathcal{H}^2 \int\frac{g}{a\mathcal{H}^2}\bigg(f_{\rm rsd}+\frac{3}{2}\Omega_{\rm dm}\bigg)^{-1}da
\,.
\end{eqnarray}

The factorised form of the source term on the right-hand-side of (\ref{deu}) suggests the second-order growing mode solution
\begin{equation}
\label{D2}
\delta_{\rm dm,p}^{(2)}(\eta,\vec x)=P(\vec x)D^{(2)}_+(\eta, \Sigma).
\end{equation}
Note that, unlike the first order solution (\ref{ef}), this second-order solution is no longer separable since the source function $\mathcal{S}(a,\Sigma)$ in Eq.~(\ref{deu}) is not in general separable. The growing mode $D^{(2)}_+$ is separable only in special cases, e.g., for the case of planar symmetry, $\Sigma=1$, or matter-dominated solutions where $\Omega_{\rm dm}$, $g$ are $f_{\rm rsd}$ are constant in time.
Nonetheless, without loss of generality we may define the local second-order growth rate as
\begin{equation}\label{sec}
f_2(\eta, \Sigma) = \frac{{D^{\prime(2)}_+}}{2\mathcal H{D^{(2)}_+}},
\end{equation}
%which is solution of the differential equation $(\ref{demais})$ with a specific shape coefficient $\Sigma$.
%
%Thus, 
where equation $(\ref{deu})$ can then be written as
\begin{equation}
 \label{D22}
4P(\vec x)\bigg(2f_2+\frac{3}{2}\Omega_{\rm dm}+g\bigg)D^{(2)}_+=\frac{(\nabla^2\mathcal{R}_c)^2}{\mathcal{H}^4} \mathcal{S}(\eta,\Sigma)
\,.
\end{equation}
Using the first-order solution $(\ref{pa})$ we can formally write the second-order particular solution as
\begin{equation}\label{part}
\delta^{(2)}_{\rm dm,p}=\frac{[2f_{\rm rsd}+3\Omega_{\rm dm}]^2}{8(4f_2+2g+3\Omega_{\rm dm})}S(a,\Sigma)(\delta_{\rm dm}^{(1)})^2.
\end{equation}
We see that a non-zero interaction, $g\neq0$, affects both the growing curvature (\ref{sabe-partic}) contributing to the source term (\ref{func}) driving the growth of structure at second order, and the second order growing mode (\ref{D2}).

\subsubsection{Homogeneous solution}

To find the homogeneous solution of the second-order evolution equation for the density contrast (\ref{demais}), we solve the second-order constraint equation $(\ref{me})$ with a constant source term, $\mathcal{R}^{(2)}$, i.e., 
\begin{equation}
4\mathcal H\delta_{\rm dm,h}'^{(2)}+2\bigg[a^2\bar\rho_{\rm dm}+2g\mathcal H^2\bigg]\delta_{\rm dm,h}^{(2)}
= {\mathcal {R}}^{(2)}_{\rm h} \,,
\end{equation}

The homogeneous solution is thus given by
\begin{equation}
\label{delta2h-general}
\delta^{(2)}_{\rm dm,h}(\eta,\vec x)=C_2(\vec x)D_+(\eta),
\end{equation}
where $D_+$ is the linear growth factor (\ref{sao}) and $C_2(\vec x)$ is given by (\ref{sao1}) replacing the first-order curvature, $\mathcal{R}^{(1)}=4\nabla^2 \mathcal{R}_c$, by the second order term, $\mathcal{R}_{\rm h}^{(2)}$.
Thus we have
\begin{equation}
\delta^{(2)}_{\rm dm,h}=\frac{\mathcal{R}^{(2)}_{\rm h}}{4\mathcal{H}^2}\bigg(f_1+\frac{3}{2}\Omega_{\rm dm}+g\bigg)^{-1} \,,
\end{equation}
where subtracting the time-dependent contribution (\ref{sabe-partic}) from full second-order curvature (\ref{biss}) gives \cite{Bruni}
\begin{align}\label{set}
\mathcal R^{(2)}_{\rm h}=4\nabla^2\bigg[\psi^{(2)}+\frac{1}{6}\nabla^2\chi^{(2)}\bigg]+32\mathcal R_c\nabla^2\mathcal R_c+12\partial^i\mathcal R_c\partial_i\mathcal R_c\nonumber \\-2[2\partial^i\nabla^2\chi^{(1)}\partial_i\mathcal R_c+\partial^i\partial_j\chi^{(1)}\partial^j\partial_i\mathcal R_c+\nabla^2\chi^{(1)}\nabla^2\mathcal R_c]\nonumber \\+\frac{1}{2}[\partial^i\partial^j\partial^k\chi^{(1)}\partial_i\partial_j\partial_k\chi^{(1)}-\partial^k\nabla^2\chi^{(1)}\partial_k\nabla^2\chi^{(1)}] \,.
\end{align}

To set the initial conditions at second order, we will introduce the primordial curvature perturbation on uniform-density hypersurfaces, $\zeta$. This gauge-invariant quantity remains constant on super-horizon scales for adiabatic perturbations~\cite{Malik:2003mv} and hence can be predicted from standard inflation models in order to set the initial conditions for the subsequent radiation and matter eras. 
We expand $\zeta$ at second order as
\begin{equation}
\label{pnG}
\zeta \approx \zeta^{(1)}+\frac{1}{2}\zeta^{(2)}=\zeta^{(1)}+\frac{3}{5}f_{NL}(\zeta^{(1)})^2\,,
\end{equation}
where we introduced the non-linearity parameter $f_{NL}$ to describe local-type primordial non-Gaussianity \cite{Wands:2010af}.

For scales well outside de horizon $(k\ll\mathcal{H}_i)$ and, therefore, at early times $(a_i\ll 1)$ we have
\begin{equation}
e^{2\zeta} = 1 - 2\bigg[\psi_i + \frac{1}{6}\nabla^2\chi_i \bigg]
%\zeta+\zeta^2=-\mathcal{R}_c-\frac{1}{2}\bigg[\psi_i^{(2)}+\frac{1}{6}\nabla^2\chi_i^{(2)}\bigg]
\,.
\end{equation}
Thus we find
\begin{equation}
\zeta^{(1)}=-\mathcal{R}_c,
\end{equation}
\begin{equation}
\psi_i^{(2)}+\frac{1}{6}\nabla\chi_i^{(2)}=-\bigg(2+\frac{6}{5}f_{NL}\bigg)\mathcal{R}_c^2.
\end{equation}
Setting initial conditions on large scales and at early times, the expression $(\ref{set})$ reduces to the large-scale limit (\ref{biss-largescale})
%\begin{equation}
%\mathcal R^{(2)}_{\rm h} = 4\nabla^2\bigg[\psi_i^{(2)}+\frac{1}{6}\nabla^2\chi_i^{(2)}\bigg]+32\mathcal R_c\nabla^2\mathcal R_c+12\partial^i\mathcal R_c\partial_i\mathcal R_c,
%\end{equation}
%and hence
\begin{equation}
\frac{\mathcal R^{(2)}_{\rm h}}{4} = 2\bigg(2-\frac{6}{5}f_{NL}\bigg)\mathcal {R}_c\nabla^2\mathcal {R}_c
 - \bigg(1+\frac{12}{5}f_{NL}\bigg)
 %(\nabla^2\mathcal R_c)^2+3
 \partial^i\mathcal R_c\partial_i\mathcal R_c.
\end{equation}

Thus the homogenous solution for the second-order density contrast (\ref{delta2h-general}) is given by
\begin{equation}\label{hom}
\delta_{\rm dm,h}^{(2)}=\frac{4}{\mathcal{H}^2}\bigg(f_{\rm rsd}+\frac{3}{2}\Omega_{\rm dm}\bigg)^{-1}\bigg[-\bigg(\frac{1}{4}+\frac{3}{5}f_{NL}\bigg)\partial ^i\mathcal{R}_c\partial_i\mathcal{R}_c+\bigg(1-\frac{3}{5}f_{NL}\bigg)\mathcal{R}_c\nabla^2\mathcal{R}_c\bigg].
\end{equation}

\subsubsection{Relativistic comoving density contrast}

The full solution for the second-order density contrast in synchronous comoving coordinates, obeying the initial constraint on large scale at early times, is thus a sum of the homogeneous solution ($\ref{hom}$) with the particular solution (\ref{part}), which gives
\begin{align}\label{secon}
\delta^{(2)}_{\rm dm}=-\frac{24}{5[2f_{\rm rsd}+3\Omega_{\rm dm}]}\bigg[\bigg(f_{NL}+\frac{5}{12}\bigg)\frac{\partial ^i\mathcal{R}_c\partial_i\mathcal{R}_c}{\mathcal{H}^2}+\bigg(f_{NL}-\frac{5}{3}\bigg)\frac{\mathcal{R}_c\nabla^2\mathcal{R}_c}{\mathcal{H}^2} \bigg] 
%\nonumber \\
%+\frac{[2f_{\rm rsd}+3\Omega_{\rm dm}]^2}{8(4f_2+2g+3\Omega_{\rm dm})}S(a,\Sigma)(\delta_{\rm dm}^{(1)})^2.
+ \frac{\mathcal{S}(a,\Sigma)} {2(4f_2+3\Omega_{\rm dm}+2g)} 
%(\delta_{\rm dm}^{(1)})^2
\left( \frac{\nabla^2\mathcal{R}_c}{\mathcal{H}^2} \right)^2
\, .
\end{align}
In this expression the first term corresponds to the large-scale/early-time part where the second-order perturbation contains information about primordial non-Gaussianity and the relativistic non-linear initial constraints. The constant $f_{NL}$ describes the level of primordial non-Gaussianity (\ref{pnG}) large scales at the end of inflation. In the absence of primordial non-Gaussianity $f_{NL}=0$. At smaller scales, well inside the Hubble horizon, the terms in the second line dominate and represent the growing non-Gaussianity due to gravitational collapse.

In $\Lambda$CDM we have $g=0$ and hence $f_{\rm rsd}=f_1$. At early times we have matter-dominated evolution, $\Omega_{\rm dm}=1$, and the linear growth function is $D_+\propto \mathcal{H}^2\propto a$ and hence the first-order growth rate (\ref{gr}) obeys $f_1=1$. The function $S(a,\Sigma)$ in Eq.~(\ref{func}) becomes a constant
%\begin{equation}
$\mathcal{S}(\Sigma) = (16/25)(5+2\Sigma)$, and 
% \,.
%\end{equation}
%
the second-order growing mode (\ref{D22}) reduces to $D_{+}^{(2)}\propto (D_{+})^2\propto a^2$. Hence the second order growth rate (\ref{sec}) $f_2=1$ (independent of the shape, $\Sigma$). 
% The second-order particular solution thus reduces to 
% %
% \begin{equation}\label{sabe-partic-g1-Omega}
% \mathcal{R}^{(2)}_{\rm p}=\frac{8a}{5H_0^2}[\partial^i\partial_j\mathcal{R}_c\partial^j\partial_i\mathcal{R}_c+(\nabla^2\mathcal{R}_c)^2],
% \end{equation}
The second-order solution for the synchronous comoving density contrast (\ref{secon}) in the early matter-dominated (Einstein-de Sitter) era is then given by \cite{bartolo}
\begin{equation}
\label{delta2EdS}
\delta^{(2)}_{\rm dm} = 
- \frac{24}{25} \bigg[\bigg(f_{NL}+\frac{5}{12}\bigg)\frac{\partial ^i\mathcal{R}_c\partial_i\mathcal{R}_c}{\mathcal{H}^2}+\bigg(f_{NL}-\frac{5}{3} \bigg) \frac{\mathcal{R}_c\nabla^2\mathcal{R}_c}{\mathcal{H}^2} \bigg] 
 + \frac{8(5+2\Sigma)(\nabla^2\mathcal{R}_c)^2}{175\mathcal{H}^4} \,.
\end{equation}
%[CORRECTED MINUS SIGN ON FIRST TERM AND CHANGED FACTOR 125 TO 175 IN LAST TERM. CHANGED $H_0$ AND $a$ BACK TO $\mathcal{H}$ AS THIS IS AN EARLY TIME SOLUTION, NOT VALID AT PRESENT TIME.]

%\subsection*{I. Model with $Q=3\alpha H\bar\rho_{\rm dm}\bar\rho_{V}/\bar\rho$}

In models (i) and (ii) the dimensionless interaction parameter $g$ is proportional to $\Omega_{\rm dm}-1$ and at early times we have $g\to0$ in the matter-dominated limit, $\Omega_{\rm dm}\to1$. Hence, as in $\Lambda$CDM, we recover the second-order Einstein-de Sitter solution (\ref{delta2EdS}) at early times, with the more general solution (\ref{secon}) with $g\neq0$ at late times when $\Omega_{\rm dm}\neq1$. 

%\subsection*{II. Model with $Q=qH\rho_V$}
%\subsubsection{Model with $g \propto 1-\Omega_{\rm dm}$ }
%In the matter-dominated era we had seen that for those models with dimensionless interaction parameter $g$ proportional to $\Omega_{\rm dm}-1$, the standard matter evolution is recovered since that $f_{\rm rsd}=f_1=1$ and $\Omega_{\rm dm}=1$. 

%\subsubsection{Model with $g=constant$}

In model (iii) the dimensionless interaction parameter $g=-\epsilon$ is non-zero at all times. Some vacuum energy is present at early times, $\Omega_{\rm dm}=1+\epsilon/3$, such that $D_{+}\propto \mathcal{H}^{-2}\propto a^{1+\epsilon}$ and hence a modified growth rate, $f_1=1+\epsilon$. The second-order source term (\ref{func}) remains a constant in this early time limit
\begin{equation}
\mathcal{S}(\Sigma) 
%=\frac{8}{(5+\epsilon)^2}[10+4\Sigma+\epsilon(1+\Sigma)]-\frac{8\epsilon(1+\Sigma)}{(5+\epsilon)(1+\epsilon)}
% = \frac{16(5+2\Sigma)}{(5+\epsilon)^2} - \frac{32\epsilon(1+\Sigma)}{(1+\epsilon)(5+\epsilon)^2}
 = \frac{16(5+2\Sigma+3\epsilon)}{(1+\epsilon)(5+\epsilon)^2}
\,,
\end{equation}
The solution for the second order density contrast is then separable and with the second-order growing mode $D_{+}^{(2)}\propto (D_{+})^2$ as in a conventional matter-dominated era, but with a modified growth rate, (\ref{sec}), $f_2=1+\epsilon$.
%The second-order growth rate and Hubble parameter, respectively, have early-time solutions $D_{+}^{(2)}\propto (D_{+}^{(1)})^2\propto a^{2(1+\epsilon)}$ and $\mathcal{H}^2=H_0^2a^{-(1+\epsilon)}$. Using again the definition $(\ref{sec})$ we find $f_2=1+\epsilon$. 
%
%The particular solution for the time-dependent curvature $(\ref{sabe-partic})$ in this model is given by
%
%\begin{equation}\label{sabe-partic-gconstant}
%\mathcal{R}^{(2)}_{\rm p}=\frac{2a^{1+\epsilon}}{H_0^2(5+\epsilon)}\bigg(4-\frac{3\epsilon}{1+\epsilon}\bigg)[\partial^i\partial_j\mathcal{R}_c\partial^j\partial_i\mathcal{R}_c+(\nabla^2\mathcal{R}_c)^2],
%\end{equation}
%
The solution (\ref{delta2EdS}) for the second-order synchronous comoving density contrast is thus 
\begin{align}
\delta^{(2)}_{\rm dm}(a,\Sigma)=-\frac{24}{5(5+\epsilon)}\bigg[\bigg(f_{NL}+\frac{5}{12}\bigg)\frac{\partial ^i\mathcal{R}_c\partial_i\mathcal{R}_c}{\mathcal{H}^2} + \bigg(f_{NL}-\frac{5}{3}\bigg)\frac{\mathcal{R}_c\nabla^2\mathcal{R}_c}{\mathcal{H}^2} \bigg]
 \nonumber \\
% + \frac{8}{(7+3\epsilon)(5+\epsilon)^2}\bigg[5+2\Sigma-\frac{2\epsilon(1+\Sigma)}{1+\epsilon}\bigg] \bigg( \frac{\nabla^2\mathcal{R}_c}{\mathcal{H}^2} \bigg)^2
+ \frac{8(5+3\epsilon+2\Sigma)}{(7+3\epsilon)(5+\epsilon)^2(1+\epsilon)} \bigg( \frac{\nabla^2\mathcal{R}_c}{\mathcal{H}^2} \bigg)^2
 \,.
\end{align}
This reduces to the standard matter-dominated (Einstein-de Sitter) second-order solution (\ref{delta2EdS}) in the limit $\epsilon\to0$.

\subsubsection{Relativistic Eulerian density contrast}

In the absence of an interaction between dark energy and dark matter, the continuity equation ($\ref{con}$) and Raychaudhuri equation ($\ref{es}$) in the synchronous comoving gauge are formally identical to the corresponding equations for the fluid dynamics in Newtonian gravity 
% since the variables in the relativistic case are identified with the variables 
in Lagrangian coordinates, i.e., comoving with the matter \cite{Bruni}. The general solution to these second-order evolution equations is thus identical to the Newtonian solution, but the relativistic solution (\ref{secon}) has a characteristic initial condition (the specific choice for the second order homogeneous solution) set by the non-linear initial relativistic constraints.

To compare our general solution (\ref{secon}) with the standard second-order solution for the density contrast in Newtonian theory, for example, we will also transform from the comoving (Lagrangian) frame to an Eulerian frame where the matter moves with respect to ``fixed'' spatial coordinates. The perturbed scalar expansion (\ref{cd}) is corresponds to the divergence of the matter 3-velocity in this frame, $\vartheta\equiv \nabla^2 v$. In relativistic perturbation theory this Eulerian frame is usually referred to as the total-matter gauge \cite{Malik:2003mv,Bertacca:2015mca}. Although the first-order density perturbation is invariant under a change of spatial gauge, at second order the density contrast transforms to \cite{Bruni,Bertacca:2015mca}
\begin{equation}
\delta_{E}^{(2)} = \delta^{(2)}_{\rm dm} - 2\partial_i\delta_{\rm dm} \int \partial^i v \, d\eta \,.
\end{equation}
Substituting in the first order results for the density contrast and velocity divergence, we find the Eulerian density
\begin{equation}
\delta_{E}^{(2)}=\delta^{(2)}_{\rm dm} + \frac{8}{(2f_{\rm rsd}+3\Omega_{\rm dm})^2} \bigg[ 1 + (2f_{\rm rsd}+3\Omega_{\rm dm}) \mathcal{H}^2 \int \frac{g}{a\mathcal H^2 (2f_{\rm rsd}+3\Omega_{\rm dm})} da \bigg] \frac{\partial^i\mathcal{R}_c\partial_i\nabla^2\mathcal{R}_c}{\mathcal{H}^4},
\end{equation}
where $\delta_{\rm dm}^{(2)}$ is given by solution ($\ref{secon}$) in synchronous comoving gauge and the second term is due to the spatial gauge transformation. 

In an early matter era, including the possibility of a non-zero interaction $g=-\epsilon$ such that $\Omega_{\rm dm}=1+(\epsilon/3)$, we can then obtain an analytic expression for the Eulerian density contrast
\begin{align}
\label{D2Euleriangconstant}
\delta_{E}^{(2)} = 
-\frac{24}{5(5+\epsilon)}\bigg[\bigg(f_{NL}+\frac{5}{12}\bigg)\frac{\partial ^i\mathcal{R}_c\partial_i\mathcal{R}_c}{\mathcal{H}^2} + \bigg(f_{NL}-\frac{5}{3}\bigg)\frac{\mathcal{R}_c\nabla^2\mathcal{R}_c}{\mathcal{H}^2} \bigg]
 \nonumber \\
+ \frac{8(5+3\epsilon+2\Sigma)}{(7+3\epsilon)(5+\epsilon)^2(1+\epsilon)} \bigg( \frac{\nabla^2\mathcal{R}_c}{\mathcal{H}^2} \bigg)^2
 + \frac{8}{(1+\epsilon)(5+\epsilon)^2} \frac{\partial^i\mathcal{R}_c\partial_i\nabla^2\mathcal{R}_c}{\mathcal{H}^4} \,.
\end{align}
We recover the early-time limit in the conventional matter-dominated limit of $\Lambda$CDM or models (i) or (ii), where $g\to0$ and $\Omega_{\rm dm}\to1$ at early time, in the limit $\epsilon\to0$.

Any separable second-order solution can be expressed in Fourier space via the convolution 
\begin{equation}
\delta_{E\vec k}^{(2)}=2\int\frac{d^3\vec{k}_1d^3\vec{k}_2}{(2\pi)^3}\delta_{D}(\vec k-\vec{k}_1-\vec{k}_2)F_2(\vec{k}_1,\vec{k}_2)\delta^{(1)}_{\vec{k}_1}\delta^{(1)}_{\vec{k}_2},
\end{equation}
with kernel
\begin{align}\label{second}
F_2(\vec{k}_1,\vec{k}_2) = F_{in}(\vec{k}_1,\vec{k}_2) + F_{nl}(\vec{k}_1,\vec{k}_2) \,,
\end{align}
where we separate two distinct contributions coming from the linearly and non-linearly growing terms.

The relativistic initial constraint including any primordial non-Gaussianity gives rise to the linearly growing term which dominates at early times (large scales)
in $\Lambda$CDM or models (i) or (ii)
\begin{align}
F_{in}(\vec{k}_1,\vec{k}_2)
= \frac{3(2f_{\rm rsd}+3\Omega_{\rm dm})}{5} \mathcal{H}^2 \bigg[ \bigg(f_{NL}+\frac{5}{12}\bigg) \frac{\vec{k}_1\cdot\vec{k}_2}{k_1^2k_2^2} + \bigg(f_{NL}-\frac{5}{3}\bigg) \frac{k_1^2+k_2^2}{2k_1^2k_2^2} \bigg]
 \,,
\end{align}
For the early matter era with $g=-\epsilon$ such that $\Omega_{\rm dm}=1+(\epsilon/3)$ this becomes
\begin{align}\label{second-Fini-gconstant}
F_{in}(\vec{k}_1,\vec{k}_2)
= \frac{3(5+\epsilon)}{5} \mathcal{H}^2 \bigg[ \bigg(f_{NL}+\frac{5}{12}\bigg) \frac{\vec{k}_1\cdot\vec{k}_2}{k_1^2k_2^2} + \bigg(f_{NL}-\frac{5}{3}\bigg) \frac{k_1^2+k_2^2}{2k_1^2k_2^2} \bigg]
 \,,
\end{align}
For $\epsilon=0$ this reduces to the conventional Einstein de-Sitter initial constraint~\cite{bartolo,bartolo1,Bruni}.

The nonlinear growth of structure due to gravitational instability and vacuum-dark matter interactions dominates at late times (small scales). For general interacting-vacuum cosmology the solution is not separable, however for the matter era solution (\ref{D2Euleriangconstant}) with $g=-\epsilon$ such that $\Omega_{\rm dm}=1+(\epsilon/3)$ we have
\begin{equation}
F_{nl}(\vec{k}_1,\vec{k}_2)
= \frac{5+3\epsilon}{(7+3\epsilon)(1+\epsilon)}
+ \frac{2}{(7+3\epsilon)(1+\epsilon)} \frac{(\vec{k}_1\cdot\vec{k}_2)^2}{k_1^2k_2^2}
+ \frac{1}{1+\epsilon}\frac{\vec{k}_1\cdot\vec{k}_2(k_1^2+k_2^2)}{2k_1^2k_2^2},
\label{Fnlepsilon}
\end{equation}
In the absence of vacuum-dark matter interactions [$\epsilon=0$ or models (i) or (ii) at early times] this reduces to the standard Newtonian kernel \cite{Bernardeau:2001qr}
\begin{equation}
F_{N}(\vec{k}_1,\vec{k}_2)
= \frac{5}{7}
+ \frac{2}{7} \frac{(\vec{k}_1\cdot\vec{k}_2)^2}{k_1^2k_2^2}
+ \frac{\vec{k}_1\cdot\vec{k}_2(k_1^2+k_2^2)}{2k_1^2k_2^2},
\end{equation}

\section{Conclusions}

In this paper we have studied the growth of density perturbations in three simple models where dark matter interacts with vacuum energy to give rise to late-time acceleration. In two of these models, including a decomposed Chaplygin gas model, the interaction vanishes at early times leading to a conventional matter-dominated (Einstein-de Sitter) cosmology. In the third model we have considered a constant dimensionless interaction rate relative to the matter density, leading to a modified matter era at early times. In all three models the interaction vanishes at late times and we recover a constant vacuum energy, driving a de Sitter expansion in the asymptotic future.

The growth of inhomogeneous perturbations of interacting dark matter is dependent upon the covariant energy-momentum transfer four-vector, $Q^\mu$. We have considered a simple interaction model where the energy-momentum transfer follows the matter four-velocity, $Q^\mu\propto u^\mu$. In this case the vacuum energy is homogeneous on spatial hypersurfaces orthogonal to the comoving worldlines and therefore 
%there are no pressure gradients in this frame \cite{Wands2009}. Thus 
the sound speed remains zero even in the presence of a non-zero matter-vacuum interaction. This means we get a simple, scale-independent growth of linear density perturbations, similar to standard cold dark matter; a non-zero sound speed would lead to a finite Jeans length, suppressing clustering on small scales.

% and we can choose a synchronous and comoving coordinate frame in which to study the growth of structure. 
We find the linearly growing mode for the first-order comoving density contrast, which in a conventional matter-dominated (EdS) era reduces to the usual linearly growing mode, $D_+\propto a$ with corresponding linear growth rate $f_1\equiv d\ln\delta/d\ln a=1$. Matter over-densities grow due to gravitational collapse and this can be enhanced by non-zero energy transfer from dark matter to the vacuum.
%, $g>0$. 
%
%We give expressions for the modified growth index (\ref{defgamma}) $\gamma=d\ln f_1/d\ln\Omega_{\rm dm}$ for each model expanding about the matter-dominated $\Omega_{\rm dm}=1$ limit. 
%The simple growth index, $\gamma=6/11$ (\ref{Peebles-index}), gives a per cent level fit to the linear mode in $\Lambda$CDM, but we find that the presence of a dark matter-vacuum interaction the error can be as large as 10-20\%, e.g., for a decomposed Chaplygin gas with dimensionless parameter $-0.1<\alpha<0.2$, as shown in figure~\ref{fig4before}.
%
For example, in the case of a non-zero energy transfer from dark matter to the vacuum even at early times, as in our model (iii) where $Q=\epsilon H\bar\rho_{\rm dm}$ we have a modified early time limit $\Omega_{\rm dm}\to1+(\epsilon/3)$ with a modified growing mode, $D_+\propto a^{1+\epsilon}$, and hence $f_1=1+\epsilon$.
Non-zero energy transfer from/to matter leads to an enhanced/suppressed matter growth rate. 
This may appear counter-intuitive, but since the vacuum is homogeneous in the comoving frame any energy transfer to the matter contributes only to the background matter density and not to the comoving density perturbation. Hence the growth rate of the local matter density contrast, $\delta_{\rm dm}=\delta\rho_{\rm dm}/\bar\rho_{\rm dm}$, is suppressed.

Energy transfer between dark matter and the vacuum also changes the usual relation between the growth rate and the velocity divergence. For interacting dark matter the linear growth rate for the matter overdensity, $f_1$, differs from the growth rate that would be inferred purely from redshift-space distortions (i.e., the peculiar velocity field) which we denote by $f_{\rm rsd}$, defined in Eq.~(\ref{deffrsd}) and related to the linear growth rate in Eq.~(\ref{frsd}). 
By contrast with the linear growth rate, the RSD distortions are enhanced by energy transfer from the vacuum to dark matter as the velocity field responds to the local gravitational potential and thus the total comoving density perturbation, not just the density contrast. 

We give expressions for the RSD index, 
\begin{equation}
\gamma = \frac{d \ln f_{\rm rsd}}{d\ln\Omega_{\rm dm}} \,,
\end{equation}
for each model by expanding about the early matter-dominated limit. The corresponding expressions for $f_{\rm rsd}\propto \Omega_{\rm dm}^{\gamma}$, give a per-cent level fit to the RSD parameter in an interacting model corresponding to the decomposed Chaplygin gas with $-0.2<\alpha<0.2$, see figure~\ref{fig4}. 
In principle independent measurements of the RSD parameter and the linear growth rate for the density contrast could reveal the effect dark matter interaction. This assumes that galaxies follow the dark matter velocity field, i.e., the role of baryons is sub-dominant in determining the peculiar velocities of galaxy. It would be interesting to develop more realistic model of a baryon+dark matter system in the presence of vacuum-dark matter interactions.

We have also found solutions for the second-order growth of the density contrast in interacting vacuum cosmologies for the first time. We identify two components in the second-order density field, Eq.~(\ref{second}), analogous to the usual second-order solutions in non-interacting $\Lambda$CDM cosmology. 

One component is a homogeneous solution, corresponding to a linearly growing density perturbation whose amplitude is second order in perturbations. This includes any primordial non-Gaussianity, e.g., originating during a period of inflation in the very early universe, as well as a term due to the initial second-order constraint for the comoving density contrast in general relativity \cite{bartolo, bartolo1, Bruni, Bruni:2014xma}, usually set to zero in Newtonian studies of structure formation \cite{Bernardeau:2001qr}. This homogeneous solution dominates in the squeezed limit or at early times, but it would also be sensitive to the effect of early radiation damping on scales below the matter-radiation equality scale $\approx 100$~Mpc \cite{Tram:2016cpy} and our analytic results do not include the effect of radiation.

The second component, which we term the particular solution, is a modification of the usual Newtonian second-order density perturbation. It leads to a growing matter bispectrum which dominates on small scales and at late times, until eventually the structure formation becomes fully nonlinear. We identify the second-order kernel or reduced bispectrum (\ref{Fnlepsilon}) and show how its shape is altered by energy transfer to or from the vacuum. This opens up the possibility of distinguishing interacting dark matter models in future through the shape of the matter bispectrum on weakly nonlinear scales (see \cite{Yamauchi:2017ibz} for related work in modified gravity). A much more challenging task for future work would be to identify dark matter-vacuum interactions in the fully nonlinear regime. Nonetheless our second order results suggest that the bispectrum, or higher order correlations in the matter density field, could in future be used to identify modifications of the standard $\Lambda$CDM scenario.

\section*{Acknowledgements}

The authors are grateful to Saulo Carneiro, Marco Bruni and Joan Sol\`a for useful discussions.
H.A.B. was partially supported by CNPq and Fapesb. 
DW was supported by STFC grant ST/N000668/1 and ST/S000550/1. 
DW is grateful to KITP, University of California Santa Barbara, for their hospitality while this paper was revised. 
This research was supported in part by the National Science Foundation under Grant No. NSF PHY-1748958.

\end{document}